\documentclass[aps,showpacs,pre,twocolumn]{revtex4}

\usepackage{psfig,amssymb,amsmath}

\def \cl{{\mathcal{C}}}
\def \ER{Erd\H{o}s-Renyi}
\def \graph{{\mathcal{G}}}
\def \order#1{{\mathcal{O}}(#1)}

\def \Es{\mathcal{B}}

\def \coordn {\gamma} 

\def \grg {\deriv{\coordn_r}{\coordn}}

\def \grh {\deriv{\coordn_r}{h}}

\def \geff {g_{\mathrm{eff}}}
\def \udof {n_F}

\def \fg {\deriv{\udof}{\coordn}}
\def \fh {\deriv{\udof}{h}}
\def \fH {\deriv{\udof}{H}}

\def \R {R}
\def \Rh {R_h}

\def \partfunc{{\mathcal{Z}}}

\newcommand{\deriv}[2] {\frac{\partial #1}{\partial #2}}

\newcommand{\Eqn}[1]{Eq.~(\ref{#1})}     
\newcommand{\Eqns}[1]{Eqns.~(\ref{#1})}     
\newcommand{\Sec}[1]{Section~\ref{#1}}     
\newcommand{\Apd}[1]{Appendix~\ref{#1}}     
\newcommand{\Fig}[1]{Fig.~\ref{#1}}     
\newcommand{\Tbl}[1]{Table~\ref{#1}}     

\begin{document}
\title{Rigidity percolation in a field~\footnote{Dedicated to Dietrich
    Stauffer, on the occasion of his 60th birthday.}}  \author{Cristian
  F.~Moukarzel~\footnote{email address: cristian@mda.cinvestav.mx}}
\affiliation{Depto.\ de F\'\i sica Aplicada, CINVESTAV del IPN,\\
  Av.~Tecnol\'ogico Km 6, 97310 M\'erida, Yucat\'an, M\'exico } 
\date{\today}
\begin{abstract}
  Rigidity Percolation with $g$ degrees of freedom per site is analyzed on
  randomly diluted \ER\ graphs with average connectivity $\coordn$, in the
  presence of a field $h$. In the $(\coordn,h)$ plane, the rigid and flexible
  phases are separated by a line of first-order transitions whose location is
  determined exactly.  This line ends at a critical point with classical
  critical exponents. Analytic expressions are given for the densities $\udof$
  of uncanceled degrees of freedom and $\coordn_r$ of redundant bonds. Upon
  crossing the coexistence line, $\coordn_r$ and $\udof$ are continuous,
  although their first derivatives are discontinuous.  We extend, for the case
  of nonzero field, a recently proposed hypothesis, namely that the density of
  uncanceled degrees of freedom is a ``free energy'' for Rigidity Percolation.
  Analytic expressions are obtained for the energy, entropy, and specific
  heat.  Some analogies with a liquid-vapor transition are discussed.\\
  Particularizing to zero field, we find that the existence of a $(g+1)$-core
  is a necessary condition for rigidity percolation with $g$ degrees of
  freedom.  At the transition point $\coordn_c$, Maxwell counting of degrees
  of freedom is exact on the rigid cluster and on the $(g+1)$-rigid-core,
  i.e.\ the average coordination of these subgraphs is exactly $2g$, although
  $\coordn_c$, the average coordination of the whole system, is smaller than
  $2g$.  $\coordn_c$ is found to converge to $2g$ for large $g$, i.e.\ in this
  limit Maxwell counting is exact globally as well.
\end{abstract}
\pacs{05.70.Fh, 64.60.Ak, 02.10.Ox}
\maketitle
\section{Introduction}
\label{sec:intro}
Scalar Percolation(SP)~\cite{SAI94,BHF96} is a paradigm for the geometric
phase transition that takes place on an initially disconnected lattice of
point-like sites, when the density $p$ of present bonds is continuously
increased. At the percolation point $p_c$ the system becomes connected on a
macroscopic scale.  This may mean that transport can happen across the system
(conductivity, fluid flow), or that the system becomes correlated on
macroscopic scales, i.e.\ ordered. It is in fact possible to describe magnetic
transitions in terms of the percolation of properly defined
clusters~\cite{CKC80,KCT81,CPC82}.  Because of the generality and simplicity
of the concepts involved, this paradigm has found multiple applications in
science~\cite{SAI94,SA94,BHF96,SN98}. In most cases, the physical variables
attached to sites are scalars, i.e.\ each site has one associated \emph{degree
  of freedom}.
\\
A generalization of this paradigm considers the case in which there is more
than one degree of freedom per site, and has been termed Rigidity
Percolation(RP)~\cite{FSP84,KWE84,FTGE85,TR85,DTTR86,TGS87,TDR99,MDC99}. In
$g$-RP, each site of a lattice has $g$ degrees of freedom, and each present
bond eliminates one relative degree of freedom.  The most commonly invoked
application of RP deals with the statics of structures.  Consider for example
the problem of bracing a \emph{framework} in three dimensions, i.e.\ rigidly
connecting a set of point-like joints by means of rotatable bars. Each joint
has three translational degrees of freedom, $g=3$, and each rotatable bar
fixes the distance between two nodes, thus providing one relative constraint.
The question of whether a given set of bars is enough to rigidize a given
structure constitutes a classical problem in applied mathematics, that of
Graph Rigidity~\cite{CS79,WI84,HC92}. In statistical physics, bonds (bars) are
randomly present with probability $p$ or absent with probability $1-p$, and
one asks for the typical rigid properties of such structures. Upon increasing
the density $p$ of present bonds, the system goes into the rigid phase,
characterized by the existence of an extensive rigidly connected cluster.  The
RP problem becomes fully equivalent to SP when $g=1$. SP has a continuous
transition in all dimensions~\cite{SAI94}. In two dimensions the RP transition
is continuous but in a different universality class~\footnote{See
  Refs.~\cite{FSP84,KWE84,JTG96,MDC99}, and also C.~Moukarzel and P.~Duxbury
  in \protect \cite{TDR99}.} than SP.  In the Mean Field(MF) limit, RP ($g>1$)
has a first-order transition~\cite{MDLF97,DJTF99}.
\\
The situation is reminiscent of the Potts model, whose MF transition is
continuous for $q=2$ and discontinuous for $q>2$~\cite{WT82,BGJT96}.  Potts
models in the presence of a field, and their relation with percolation models,
have been studied recently because of possible links with the deconfining
transition in QCD~\cite{KSSC02,SS02}. In the presence of a field, and for
large enough $q$, the Potts model has a line of first-order transitions ending
at a critical point~\cite{KST00,FSC02}.  This critical point appears to always
be in the Ising universality class.
\\
Nonzero field values have been considered in scalar percolation studies
previously~\cite{KP76,GSGS77,SS77,RSKP78,NRS79,EP80,RSKL80}. A field may be
introduced in percolation by allowing for the existence of ``ghost bonds''
which are present with probability $h$ and connect sites directly to a solid
background (or to ``infinity'').  However, for any nonzero field there is no
SP transition.  RP is somewhat more interesting, as we will find out.  In this
work, RP is studied on diluted random graphs of the \ER\ type, with average
connectivity $\coordn$, in the presence of a field $h$.  Unlike SP, mean-field
RP has, in the presence of a field, a line of first-order phase transitions.
This line ends at a critical point with classical critical indices:
$\alpha=0$, $\beta=1/2$, $\gamma=1$ and $\delta=3$.
\\
Some of the analysis in zero field is relevant for the related problem of
Bootstrap Percolation (BP)~\cite{CLRB79,AB91,KSA93,MMD99}, also known as
$k$-core~\cite{BT84} in the field of Graph Theory~\cite{LS91,MA96b,PSWS96}. In
$k$-BP~\cite{CLRB79}, all sites with less than $k$ neighbors are iteratively
culled. What remains, if something, is the $k$-core~\cite{BT84}; a subgraph
were all sites have $k$ or more neighbors.  In $g$-RP a site needs at least
$g$ bonds in order to be attached to a rigid cluster.  Thus the ``infinite''
rigid cluster is a subset of the $g$-core.  In \Sec{sec:stateeqn} we will see
that an even stronger condition exists for rigidity: \hbox{$g$-RP} requires
the existence of a \hbox{$(g+1)$-core}~\footnote{{}However, the rigid cluster
  is not a subset of the \hbox{$(g+1)$-core}}.
\\
Our approach starts by deriving an equation of state for the ``order
parameter'' $R(\coordn,h)$, the probability that a randomly chosen site
belongs to the rigid cluster, as a function of $\coordn$, the average number
of bonds impinging on a site, and $h=\coordn H$, where $H$ is the applied
field.  We will call $H$ or $h$ indistinctly the ``field'' variable.  The
equation of state, as is customary in these cases, is found to accept multiple
solutions. Stability analysis is not enough to single out a unique solution.
Of central importance in order to lift this multiplicity are
$\coordn_r(\coordn,h)$, the average number of ``redundant'' bonds per site
(see later), and $\udof$, the average number of uncanceled degrees of freedom
per site.  Their relevance resides in the fact that they \emph{must} be
continuous functions of $\coordn$. Requiring that $\udof$ (or, equivalently,
$\coordn_r$) be continuous is enough to identify the physically correct
solution.
\\
This work is similar in spirit to previous treatments of RP in zero field on
Bethe lattices~\cite{DJTF99}, i.e\ networks where each site has exactly $z$
randomly chosen neighbors.  We consider the effect of an external field, and
particularize to random graphs of the \ER\ type~\cite{BR01}.  The introduction
of a field appears to be much more tractable analytically on \ER\ graphs than
on Bethe lattices, and this is the main reason why most of the results
presented here are for \ER\ graphs.  For these, we are able to derive analytic
expression for the densities $\udof$ of uncanceled degrees of freedom and
$\coordn_r$ of redundant bonds for arbitrary values of the field.  However in
zero field calculations are also straightforward for Bethe lattices, and allow
us to obtain, in an entirely analytic fashion, some of the results obtained by
numerical integration in Ref.~\cite{DJTF99}.
\\
It is known that SP and Potts models are particular cases of a more general
Fortuin-Kasteleyn Random-Cluster(RC) model, defined by a continuous parameter
$q$~\cite{KFP69,FKR72}. SP can be obtained as the $q\to 1$ limit of this
Random Cluster model and, in this limit, the logarithm of the partition
function coincides with the average number of connected clusters.  It seems
possible that a similar mapping might exist for RP as well, although it has
not been found up to now.  However it has been proposed~\cite{JTG96,DJTF99}
that the number of uncanceled degrees of freedom $\udof$ is a good ``free
energy'' candidate for RP in zero field. When $g=1$, each connected cluster
has one uncanceled degree of freedom, so both definitions coincide in this
limit. We explore on this idea further in this work.
\\
It is possible to establish a pedagogical parallel between the RP transition
and a Condensation transition. One identifies the coordination parameter
$\coordn$ with an inverse temperature $\beta$; the negative of the order
parameter $R$ plays the role of the fluid volume $V$ (it is also possible to
identify $R$ with $\rho$, the fluid density)~\footnote{ For a discussion see
  for example \protect \cite{SI87}.}, and the field \hbox{$H = h/\coordn$} is
the fluid pressure $P$. Within this analogy, it results natural to argue that
requiring the continuity of $\udof$ in order to identify the physically
correct solution $R$ in Rigidity Percolation is equivalent to requiring the
continuity of the free energy (giving rise to the Maxwell construction) in the
Statistical Mechanics treatment of condensation transitions at the MF
level~\cite{SI87}.
\\
It is shown in this work that the idea of identifying $\udof$ with a free
energy for RP leads to consistent results in the presence of a field as well.
In \Sec{sec:udof-free-energy} it is shown that: 1) the order parameter $R$ is
obtained as a derivative of the free energy with respect to the field, 2) the
condition of stability (\Sec{sec:stability-analysis}) may be related to the
positivity of a suitable second derivative of the free energy, and, 3) the
continuity of $\udof$ can be cast exactly in the form of Maxwell's Rule of
Equal Areas on the RP equivalent of a $P$-$V$ diagram (which is the $H$-$R$
diagram, see \Fig{fig:Hvsr.g.2}).
\\
This work is organized as follows: The system under consideration is defined
in \Sec{sec:rand-dilut-graphs}, and its equation of state is derived in
\Sec{sec:stateeqn}, establishing its connections with Bootstrap Percolation.
In \Sec{sec:ghostfield} the field is introduced. \Sec{sec:solutions} starts
the analysis of solutions of the general equation of state, discussing
stability and the existence of a critical point.  The concept of redundant
constraints is introduced in \Sec{sec:definitions} and their density is
calculated in \Sec{sec:calculation-coordn_r}. This result is used in
\Sec{sec:locat-trans-point} to determine the value $\coordn_c$ where the
first-order transition takes place. In \Sec{sec:case-zero-field}, the counting
of constraints is done on the $(g+1)$-rigid core in zero field, both for \ER\ 
graphs and for Bethe lattices. \Sec{sec:udof-free-energy} discusses several
consequences of identifying the density of uncanceled degrees of freedom with
a free energy, and \Sec{sec:discussion} contains a discussion of the results.
\section{Setup}
\label{sec:setup}
\subsection{Randomly diluted graphs}
\label{sec:rand-dilut-graphs}
We consider graphs made of $N$ sites (or ``nodes'') where each of the
$N(N-1)/2$ pairs of distinct sites is connected by a bond (or ``edge'')
independently with probability $p$.  This defines~\cite{BR01} an \ER\ graph
with average coordination number $\coordn=p(N-1)$. In this work $\coordn$ will
be taken to be of order one. For large $N$, a site of this graph is connected
to $k$ other sites with poissonian probability $P_k(\coordn)=e^{-\coordn}
\coordn^k/k!  $.
\\
As appropriate for Rigidity Percolation, each node of this graph is regarded
as a ``body'' with $g$ degrees of freedom. For example, rigid bodies in $d$
dimensions have $d$ translational degrees of freedom plus $d(d-1)/2$
rotational degrees of freedom, for a total of $g=d(d+1)/2$. In this work, $g$
is taken to be an arbitrary integer. Each present bond represents one
\emph{constraint}, that is, removes one degree of freedom. In order to have a
physical representation in mind, a bond can be though of as a rotatable bar
that fixes the distance between two arbitrary points belonging to the bodies
it connects~\cite{MA96,MDC99}. For the class of graphs that we consider, at
most one bond is allowed for each pair of nodes.
\\
As the number of present bonds grows, parts of this graph will become
``rigidly connected''.  Rigid connectivity of a subgraph means (this is our
definition) that the total number of degrees of freedom in the system cannot
be further reduced by additional bonds connected between nodes of this
subgraph. Such subgraphs are customarily called ``rigid clusters''. A rigid
cluster has no internal degrees of freedom left. When the coordination
$\coordn$ is large enough, the largest rigid cluster encloses a finite
fraction of the system's sites, and rigidity is said to percolate. For low
$\coordn$ the system is in the flexible, or floppy, phase, and there are no
extensive rigid clusters.
\\
A rough estimate of the threshold for the appearance of an extensive rigid
cluster can be obtained by equating the average number of constraints per
site, which is $\coordn/2$, to $g$. This results in the so called Maxwell
estimate~\cite{TGS87,JTG95,JTG96,DJTF99}, $\coordn_c^{Maxwell}=2g$. This
estimate becomes exact in the $g>>1$ limit (\Sec{sec:case-zero-field}).
\subsection{The Equation of State in zero field}
\label{sec:stateeqn}
Let us now define $\cl$ to be the largest rigid cluster in the graph, if one
exists. A site in $\cl$ will be said to be a ``rigid site''.  Let $R$ be the
probability that a randomly chosen node $i$ be in $\cl$.  In order for $i$ to
be in $\cl$ it is necessary and sufficient that it be connected to $g$ or more
other nodes $j$ in $\cl$.  Our definition of rigidity is thus \emph{recursive}
at this stage.
\\
A site $j$ is rigid and connected to $i$ with probability $pR$.  Therefore $i$
has exactly $k$ rigid neighbors with probability \hbox{${N-1 \choose k} (pR)^k
  (1-pR)^{N-1-k}$}.  For $N$ large and defining \hbox{$x=\coordn R$}, the
probability $P_k$ to have exactly $k$ rigid neighbors may be written as
\begin{equation}
P_k(x)=e^{-x}x^k/k!,
\label{eq:Pk}
\end{equation}
showing that the number $k$ of \emph{rigid neighbors} of a randomly chosen
site is a poissonian variable with average \hbox{$x=\coordn R
  =\sum_{k=0}^{\infty}k P_k(x)$}.  Since a site must have $g$ or more rigid
neighbors in order to be itself rigid, we conclude that $R$ satisfies the
self-consistent equation:
\begin{eqnarray}
R &=& \sum_{k=g}^{\infty} P_k(\coordn R) 
= G_{g}(\coordn R),
\label{eq:bd3}
\end{eqnarray}
where we have defined
\begin{equation}
G_{m}(x)= \sum_{k=m}^{\infty} P_k(x).
\label{eq:F}
\end{equation}
\subsubsection*{Bootstrap Percolation, or $k$-core}
Let us now briefly discuss the related problem of Bootstrap
Percolation~\cite{CLRB79} or $k$-core\cite{BT84}.  We want to assess the
probability $BP(g+1)$ that a randomly chosen site $i$ be part of the
$(g+1)$-core. Assume this is the case.  By following one of its links, a
neighbor $j$ is reached. $j$ must have at least $g$ other neighbors. We call
these neighbors of $j$, other than the site $i$ from which we arrived at it,
the ``outgoing'' neighbors of $j$. Each of these in turn must have $g$ or more
outgoing neighbors, and so on. More formally, let us define the property of
$g$-outgoing-connectedness ($g$-OC) in the following (recursive) way: A site
is $g$-OC if $g$ or more of its outgoing neighbors also are.
\\
On graphs of the \ER\ type, the probability to have $k$ outgoing neighbors is
the same as that to have $k$ neighbors altogether, i.e.\ $P_k(\coordn)$
(\Eqn{eq:Pk}). This is so because links are independently present. Letting
$\hat R$ be the probability that a site be $g$-OC, by the same reasoning as in
the previous Section we conclude that a random site is connected to exactly
$k$ outgoing $g$-OC neighbors with probability $P_k(\coordn \hat R)$. Thus
$\hat R$ satisfies the same self-consistent equation (\ref{eq:bd3}) as $R$
does in the RP problem with $g$ degrees of freedom, i.e.\ 
\begin{eqnarray}
\hat R = G_{g}(\coordn \hat R).
\label{eq:bd4}
\end{eqnarray}
Notice that, by requiring that each site in a tree have $g$ or more
\hbox{$g$-OC } neighbors we ensure that all sites, except perhaps for the top
one, have $(g+1)$ or more neighbors.  The probability $BP(g+1)$ that the top
site itself has $g+1$ or more neighbors (which are $g$-OC) is then given by
\begin{equation}
BP(g+1) = G_{g+1}(\coordn \hat R) = \hat R - P_g(\coordn \hat R).
\label{eq:bp}
\end{equation}
This expression gives the density of the $(g+1)$-BP infinite cluster, (or
\hbox{$(g+1)$-core}) at the point where it first
appears~\cite{CLRB79,LS91,PSWS96}. So one must first solve (\ref{eq:bd4}) in
order to obtain $\hat R$ as a function of $\coordn$, and then use
(\ref{eq:bp}) to find $BP(g+1)$. Numerical results~\footnote{C.~Moukarzel, to
  be published.} show that (\ref{eq:bp}) is exact for large $N$.
\\
We see that the $(g+1)$-core density $PB(g+1)$ is somewhat smaller than $\hat
R$ whenever $\hat R>0$, while $\hat R$ in turn satisfies the same equation as
the density $R$ of rigid sites in $g$-RP. Later in \Sec{sec:case-zero-field}
we will see that in zero field, whenever there is a $g$-rigid cluster, it
contains as a subset the $(g+1)$-core.
\subsection{Equation of State in the presence of a ghost field}
\label{sec:ghostfield}
We now introduce a ``ghost field'' $H$ that couples to the order parameter
$R$, in the following way~\cite{EP80}: in addition to the ``normal'' bonds of
our graph, a number $N \coordn H = Nh$ of ``ghost-bonds'' are assumed to
exist, each connecting one site to a unique rigid background~\footnote{{}This
  rigid background replaces the ``ghost-site'' in SP. For an alternative way
  to introduce a field, see \protect \cite{KP76}}. Each ghost bond provides
\emph{one} constraint, i.e.\ removes one degree of freedom.  Multiple
occupation is allowed, so that a random site is connected to the background by
$n$ ghost bonds with poissonian probability $P_n(h)=e^{-h} h^n/n!  $.
\\
It is easy to see that all extensive rigid clusters are rigidly connected to
the background. From now on, a site is said to be rigid if it is rigidly
connected to the background, either directly through ghost bonds, or
indirectly through rigid neighbors.
\\
If a site has $n \geq g$ ghost edges connecting it to the background, then it
is rigid. Otherwise if $n<g$, it is rigid if it has, in addition to these $n$
ghost bonds, $(g-n)$ or more rigid neighbors. Thus using (\ref{eq:Pk}) we may
write, in the presence of a field $h =H \coordn$,
\begin{eqnarray}
  R &=& 
\sum_{n=g}^{\infty} P_n(h) + 
\sum_{n=0}^{g-1} P_n(h) \sum_{k=g-n}^{\infty} P_k(\coordn R) 
\label{eq:h1}
\end{eqnarray}
After a simple resummation, this expression reads
\begin{eqnarray}
  R =G_{g}(\coordn R+h),
\label{eq:h2}
\end{eqnarray}
with $G_{g}$ given by (\ref{eq:F}). \Eqn{eq:h2} generalizes (\ref{eq:bd3}) in
the presence of a field, and is the equation of state for our problem.
\\
The simplicity of (\ref{eq:h2}) is one of the reasons leading us to study this
particular (Poissonian) definition of the field. Other field definitions, like
for example assuming that each site is rigidly attached to the background with
probability $h$, or other random graph structures like Bethe
lattices~\cite{DJTF99} are also tractable with the methods used here, but the
algebra becomes more complicated.
\\
Clearly $y=\coordn(H+R)$ plays the role of a ``Weiss field'' in the MF
equation for a ferromagnet. By analogy we may thus identify $\coordn$ as the
inverse temperature, and $H$ as the magnetic field.  It is illustrative to
take notice of the similarities between (\ref{eq:h2}) and other MF equations.
For the Potts model~\cite{CMTE00},
\begin{equation}
m =\Theta_{q}(\beta m + h),
\label{eq:pottsmf}
\end{equation}
where $\beta$ is the inverse temperature, $m$ is the magnetization, $h=\beta
H$ is the external field, and \hbox{$\Theta_{q}(y)=(e^y-1)/(e^y+(q-1))$}.
\\
Notice that when $g=1$ (Scalar Percolation), (\ref{eq:h2}) gives
\hbox{$G_{1}(y)=1-e^{-y}$}, the same as (\ref{eq:pottsmf}) for $q=1$.  This is
of course just a consequence of the known equivalence between Scalar
Percolation and the $q\to 1$ limit of the Potts model\cite{KFP69,FKR72}.
\\
However, the archetypal example of a first-order transition with a
two-parameter phase space is the condensation transition~\cite{SI87},
described at the Mean-Field level by the Van der Waals equation.  The reduced
form of the Van der Waals equation can be written as
\begin{equation}
\lambda =\Gamma(\beta (\lambda + P)),
\label{eq:vdwmf}
\end{equation}
where \hbox{$\Gamma(y)=27y^2/(y+8)^2$}, $P$ is the pressure, $\beta$ is the
inverse temperature and $\lambda=3 \rho^2$ with $\rho$ the number density.
$\Gamma$ has the same general features as $G$ and $\Theta$, namely it starts
at zero, grows sharply, and saturates for large values of its argument.
Therefore all three systems give rise to the same phenomenology, including of
course sharing the same (classical) critical indices at their critical points.
\\
Within an analogy with a condensation transition, in the RP problem $\coordn$
plays the role of an inverse temperature, while $H =h/\coordn$ plays the role
of a pressure.
\section{Analysis of solutions}
\label{sec:solutions}
\begin{figure}[h]
\centerline{\psfig{figure=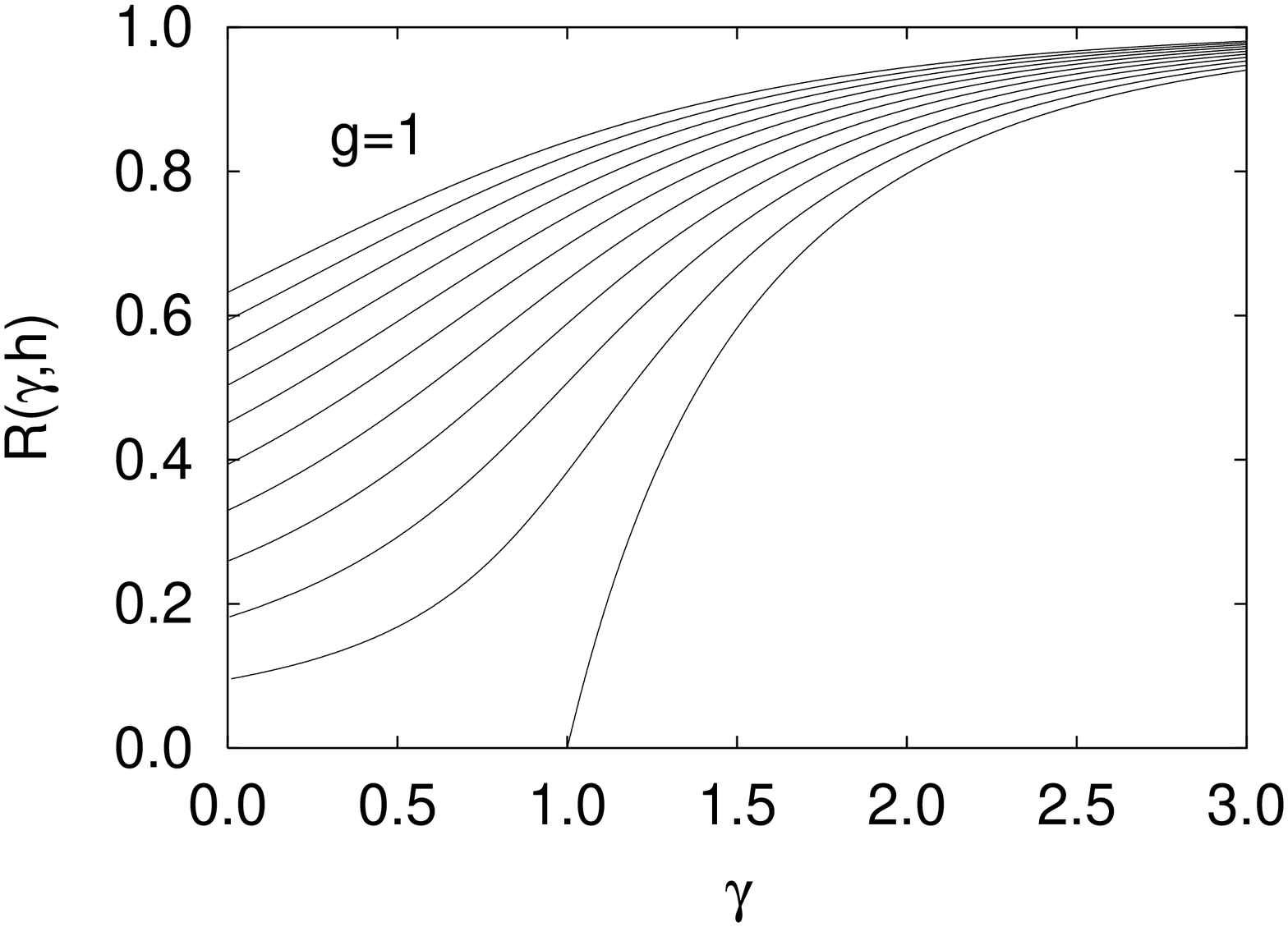,width=8cm,angle=270}}
\centerline{\psfig{figure=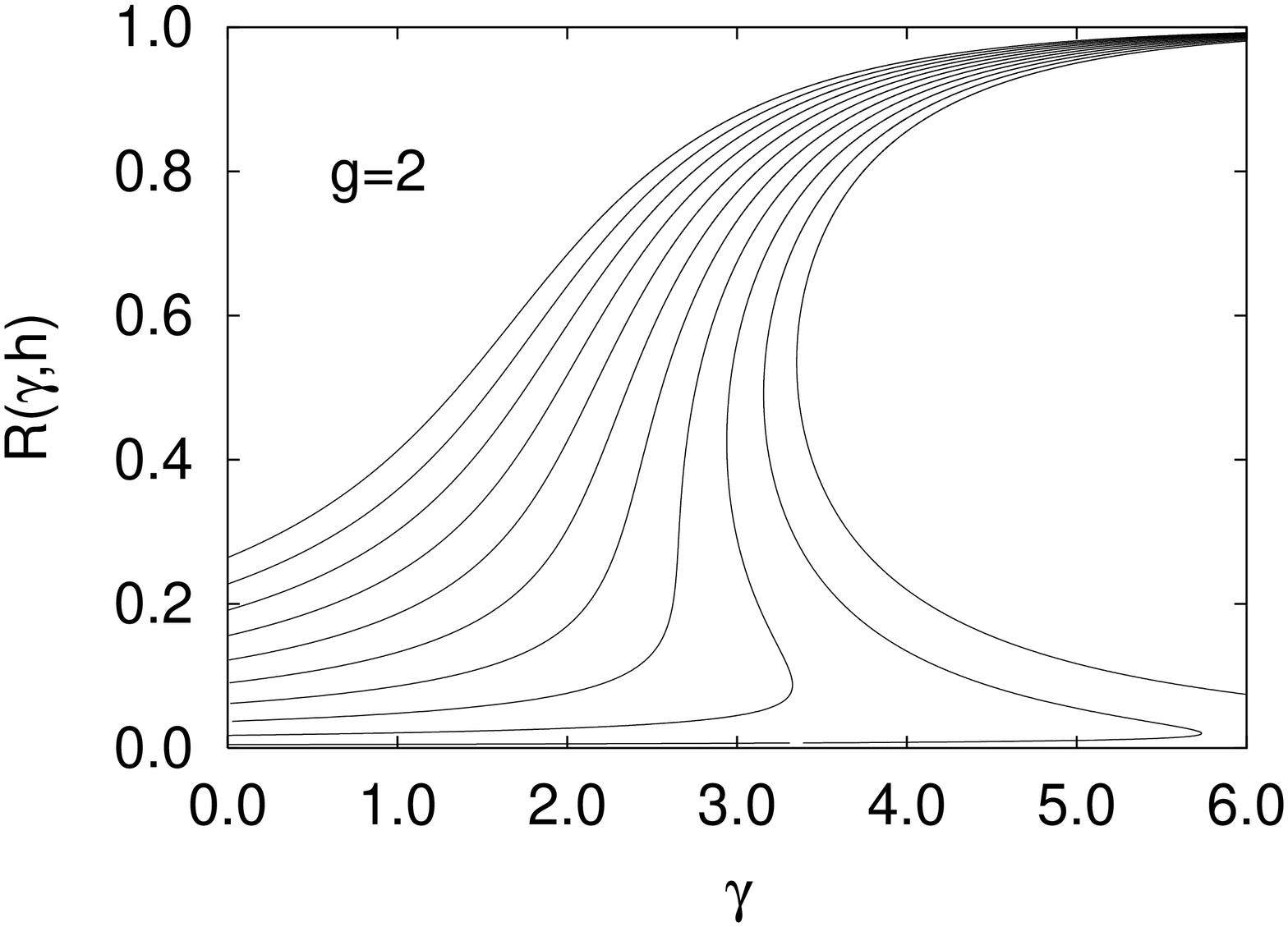,width=8cm,angle=270}}
\caption{{}The density $R$ of ``rigid sites'' as given by \Eqn{eq:h2}, for
  $g=1$ (scalar percolation, top) and $g=2$ (rigidity percolation, bottom).
  The field takes the values (from right to left):
  $h=0.0,0.1,0.20,\ldots,1.0$.  For $g=2$ the critical field is $h_c=0.287$.}
\label{fig:g1g2}
\end{figure}

In this Section, the solutions of our equation of state (\ref{eq:h2}) are
discussed.  In order to obtain $R(\coordn,h)$ numerically for given $\coordn$
and $h$, one might for example iterate (\ref{eq:h2}) until a desired numerical
accuracy is reached.  This procedure was used in Refs~\cite{MDLF97,DJTF99}.
However in this work the following alternative procedure was preferred: given
$h>0$ fixed, and for a sequence of values of $y>h$, evaluate
\begin{equation} \left \{
\begin{array}{rcl}
R &=& G_{g}(y)
\\ \\
\coordn &=&  (y-h) /G_{g}(y)
\end{array} \right .
\label{eq:bd5}
\end{equation}
thus defining $R(\coordn,h)$ implicitly. In this way one obtains the results
displayed in \Fig{fig:g1g2}. This procedure allows us to obtain all solutions
of (\ref{eq:h2}), while the iterative procedure mentioned above only provides
the stable branches.
\\
In zero field, $R=0$ is a solution $\forall \coordn$ and $\forall g$.  If
$\coordn$ is large enough, nontrivial solutions $R>0$ exist as well. If $h>0$
one has on the other hand that $R>0$ $\forall \coordn$.
\\
In the SP case ($g=1$), $G_{1}(x)=1-\exp(-x)$. Thus if $h=0$ a non-zero
solution for $R$ first appears at $\coordn_c=1$.  Above $\coordn_c$ one has $R
\approx (\coordn-1)$. In this case there is a \emph{continuous} transition at
$\coordn=1$, a well known result for scalar percolation on random
graphs~\cite{BR01}.  If $h>0$ there is no transition (\Fig{fig:g1g2}).
\\
A richer behavior is found for the RP case ($g>1$), as shown in \Fig{fig:g1g2}
for $g=2$. In this case $\coordn$, as given by \Eqn{eq:bd5}, may no longer be
a monotonous function of $y$, and, for this reason, $R$ becomes a multivalued
function of $\coordn$.  This allows for the existence of a first-order
transition. The physical considerations which lead to the identification of
the correct transition point will be discussed in
Sections~\ref{sec:stability-analysis} and~\ref{sec:locat-trans-point}.
\subsection{Spinodal Points}
\label{sec:spinodal-points}
\begin{figure}[h]
\centerline{\psfig{figure=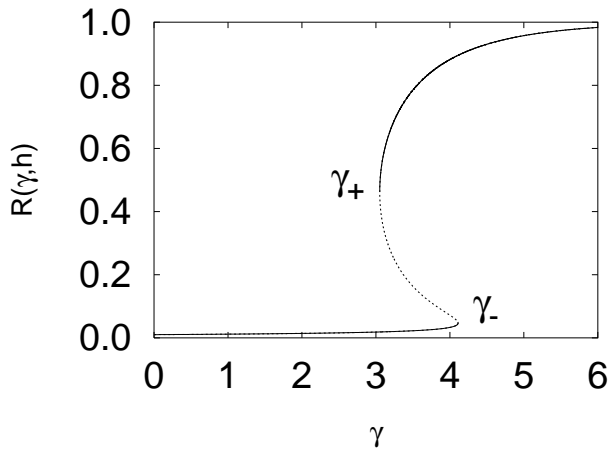,width=9cm,angle=270}
}
\caption{{} For all $g>1$, the order parameter $R$ as given by \protect
  (\ref{eq:bd5}) becomes multivalued for small values of the field $h$.  The
  branch joining the spinodals $\coordn_{\pm}$ (dashed) has $dR/d\coordn<0$
  and is thus unstable.  The uppermost and lowermost branches (solid) are
  stable.  In this example, $g=2$ and $h=0.15$. For zero field the spinodal
  $\coordn_{-}$ goes to infinity, and the branch $0\leq \coordn \leq
  \coordn_{-}$ collapses onto the $R=0$ solution, which is stable for all
  $\coordn$.}
\label{fig:stability}
\end{figure}

The condition that $\coordn$ as given by \Eqn{eq:bd5} be stationary in $x$
reads
\begin{equation}
G_{g}-x \deriv{G_{g}}{x} =0,
\label{eq:maximum}
\end{equation}
and has two solutions $x_{\pm}^s(g,h)$ for all $h$ smaller than a critical
value $h^{*}(g)$. These two solutions in turn define $\coordn_{\pm}^s$, which
are turning points for $R(\coordn,h)$ (See \Fig{fig:stability}). In the
interval $\coordn_{+}^s<\coordn<\coordn_{-}^s$, $R(\coordn,h)$ has three
solutions, two of which are stable as we show next. Therefore a
\emph{discontinuous} $\coordn$-driven transition takes place when $h<
h^{*}(g)$.  Right at $h=h^{*}$ the transition becomes continuous (See
\Sec{sec:criticalpoint}), while for $h>h^{*}(g)$, $R$ is a smooth function of
$\coordn$ and no transition occurs. \\
The thresholds $\coordn_{\pm}^s$ can be identified as \emph{spinodal-points},
as first discussed in~\cite{DJTF99}. The true rigidity percolation transition
happens at a value $\coordn_c(g,h)$ that lies in between the spinodals, and
which we analytically determine later in \Sec{sec:locat-trans-point}.
\subsection{Critical Point}
\label{sec:criticalpoint}
When $h$ takes a critical value $h^*$, the spinodals $x_{\pm}^s$ coalesce onto
an inflexion point where $\partial G_{g}(x+h)/\partial x$ equals $G_{g}/x$
(since \Eqn{eq:maximum} holds).  The \emph{critical point}
$\{h^*,\coordn^*,R^*\}$ is thus defined by
\begin{subequations}
\begin{eqnarray}
\left . \frac{\partial^2 G_{g}}{\partial x^2} \right |_{x^*,h^*}
&=& 0 
\label{eq:critconda}
\\ \nonumber 
\\
\left . \deriv{G_{g}}{x} \right |_{x^*,h^*}
&=& \frac{G_{g}(x^*+h^*)}{x^*} = \frac{1}{\coordn^*}.
\label{eq:critcondb}
\end{eqnarray}
\end{subequations}
For $h>h^*$ there is no phase transition.  Using \Eqn{eq:h2}, and defining
$y=x+h$, these two equations can be solved exactly. The critical point turns
out to be
\begin{equation} 
\left \{
\begin{array}{rcl}
 y^* &=& g-1 \\ \\
\coordn^* &=& \{ P_{g-1}(g-1)\}^{-1} = e^{g-1} (g-1)!/(g-1)^{g-1}\\ \\
R^* &=& G_{g}(g-1) = e^{1-g}\sum_{k=g}^\infty (g-1)^k/k! \\ \\
h^* &=& y^* - x^* = g-1 - \coordn^* R^* 
\end{array}
\right .
\label{eq:critpoint}
\end{equation}
For the scalar case ($g=1$) this gives $h^*=R^*=0$ and $\coordn^*=1$ in
accordance with the fact~\cite{BR01} that for zero field there is a continuous
transition at $\coordn_c=1$.
\\
For $g=2$ one finds
\begin{equation} 
\left \{
\begin{array}{rcl}
\coordn^* &=& e \\ \\
R^* &=& (e-2)/e \\ \\
h^* &=& 3- e.
\end{array}
\right .
\label{eq:critpointg2}
\end{equation}
For large $g$, and approximating $n! \approx (n/e)^n (2 \pi n)^{1/2}$ one sees
that $\coordn^* \propto g^{1/2}$, while $h^{*} \sim g$ and thus $H^* \sim
g^{1/2}$. This means that in the limit $g>>1$ most constraints are
field-constraints at the critical point.
\subsubsection{Critical Indices}
\label{sec:critical-indices}
The RP transition on random graphs is similar to other MF transitions with
first-order lines, as discussed in \Sec{sec:ghostfield}. Thus it can be
concluded that RP must have classical critical indices: $\beta=1/2$,
$\delta=3$ and $\gamma=1$. For completeness we show that this is indeed the
case by deriving the critical indices briefly in \Apd{sec:deriv-crit-indic}.
\subsection{Stability analysis}
\label{sec:stability-analysis}
Stability can be analyzed if \Eqn{eq:h2} is interpreted as a recursion
relation,
\begin{eqnarray}
R_{t+1}  = G (\coordn R_t + h),\qquad t=0,1,\ldots,\infty,
\label{eq:rec}
\end{eqnarray}
for fixed $\coordn$ and $h$. Assume $R(\coordn,h)$ is a fixed point of
\Eqn{eq:rec}, i.e.\ $R=G(\coordn R,h)$. This fixed point is stable if
\begin{equation}
\left |
\left .
\deriv{G(\coordn R +h)}{R}
\right |_{\coordn,h}
\right | = \left | \left . \coordn \deriv{G(x+h)}{x} \right
  |_{x=\coordn       R} \right | <1.
\label{eq:stability0}
\end{equation}
Since $G'(y)=P_{g-1}(y)$ we conclude that the stability condition reads
\begin{equation}
\coordn P_{g-1}(y) <1,
\label{eq:stability1}
\end{equation}
with $y=x+h$ as defined previously.
\\
A more useful form results if one notices that, if $R>0$, \Eqn{eq:stability0}
is equivalent to requiring that $dR/d\coordn>0$.  In fact,
\begin{eqnarray}
\frac{dR}{d\coordn} &=& 
\frac{dG}{dx} \frac{dx}{d\coordn} =
G' \frac{d}{d\coordn}
\left ( \coordn R \right) =  G'
\left ( R + \coordn \frac{dR}{d\coordn} \right ) \nonumber \\
&& \Rightarrow \quad \frac{dR}{d\coordn} ( 1-\coordn G' )=  G'R
\label{eq:stability2}
\end{eqnarray}
and since $G'>0$ $\forall x>0$, we conclude that a nontrivial fixed point
$R(\coordn,h)>0$ of \Eqn{eq:rec} is stable if and only if $dR/d\coordn>0$.
\\
This condition has a simple physical meaning: increasing the average
connectivity $\coordn$ should not decrease the rigid density $R$. A similar
stability condition holds for fluids, namely that the coefficient of thermal
expansion be positive.
\subsubsection{Scalar Percolation ($g=1$)}
In zero field, $R=0$ is a solution of \Eqn{eq:h2}, thus a fixed point of
\Eqn{eq:rec}, $\forall g$.  If $g=1$, stability (\Eqn{eq:stability1}) requires
that $\coordn P_0 = \coordn e^{-\coordn R}<1$. Thus the trivial solution $R=0$
becomes unstable for $\coordn>1$, where the nontrivial solution (stable
because $dR/d\coordn>0$. See \Fig{fig:g1g2}) first appears. This situation is
typical of continuous transitions; the ordered solution appears exactly at the
point where the paramagnetic solution becomes unstable.
\\
When $h>0$ there is only one solution for \Eqn{eq:h2}, and it is stable for
all $\coordn$ since $dR/d\coordn>0$ (\Eqn{eq:stability2}).
\subsubsection{Rigidity Percolation ($g>1$)}
For $g>1$ the situation is more interesting, as several stable solutions of
\Eqn{eq:h2} can coexist.  If $h<h^*$, \Eqn{eq:h2} has three solutions (see
\Fig{fig:stability}) in the range $\coordn_{-}^s>\coordn>\coordn_{+}^s$. The
branch joining the two spinodal points can be discarded since it is unstable
(because $dR/d\coordn<0$).  However the other two branches (solid lines in
\Fig{fig:stability}) are stable so that there is coexistence of two stable
solutions for $\coordn_{-}^s>\coordn>\coordn_{+}^s$. Thus the system undergoes
a $\coordn$-driven \emph{first-order} transition somewhere between the
spinodals. The precise point $\coordn_c$ at which the system switches from one
stable solution to the other is uniquely defined by a continuity requirement,
as we elaborate later in \Sec{sec:locat-trans-point}.
\\
The $h=0$ case is similar, however in this case the lowest stable branch
collapses onto the trivial solution $R=0$, while the spinodal $\coordn_{-}$
goes to infinity.
\section{Zero Modes, Redundant and Overconstrained bonds}
\label{sec:gammac}
In this Section the counting of uncanceled degrees of freedom and the useful notions of
redundant and overconstrained bonds are discussed. These will be of central
importance in our subsequent treatment of the RP problem.
\subsection{Definitions}
\label{sec:definitions}
\begin{figure}[h]
\centerline{
\psfig{figure=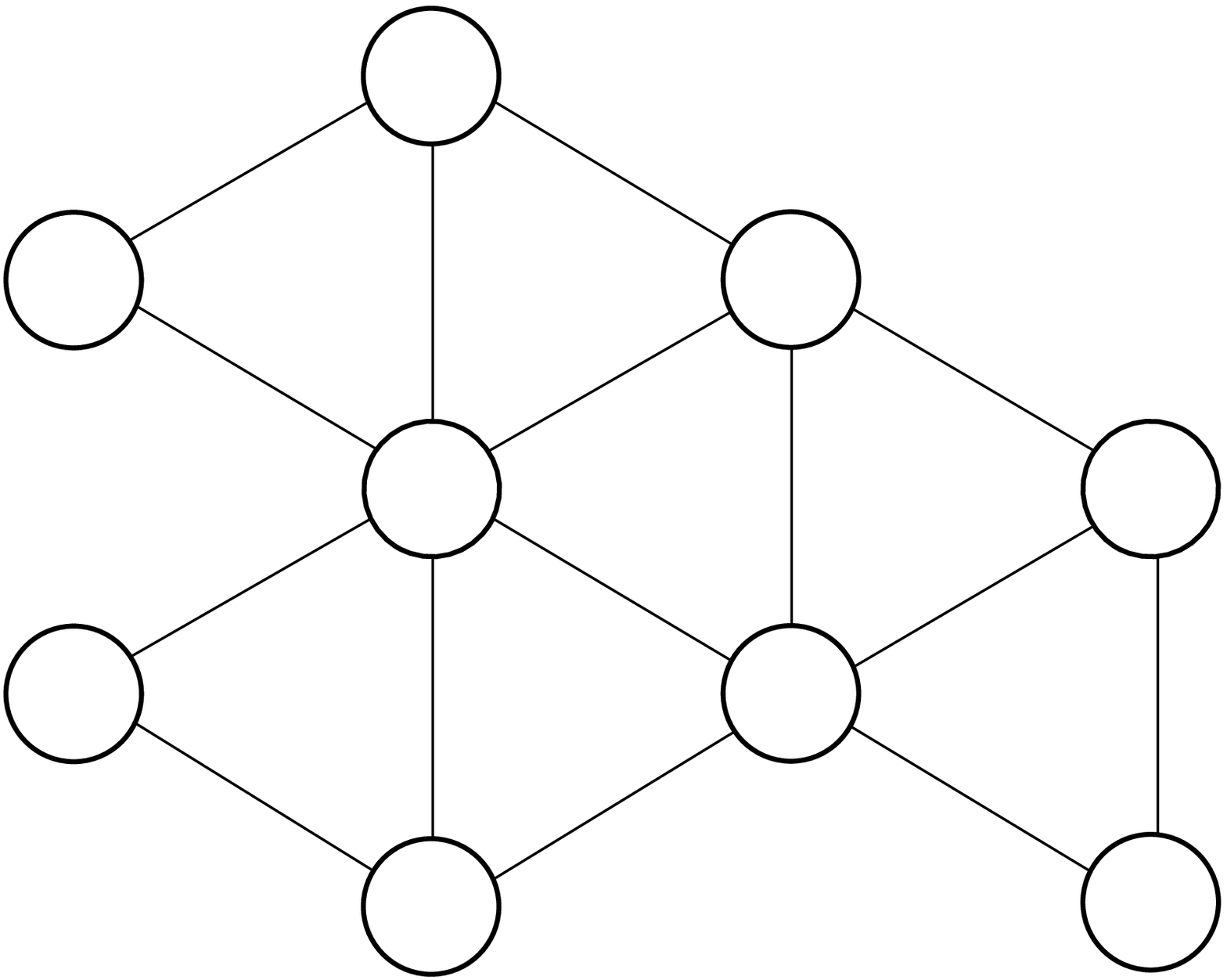,width=4.cm,angle=0}
\psfig{figure=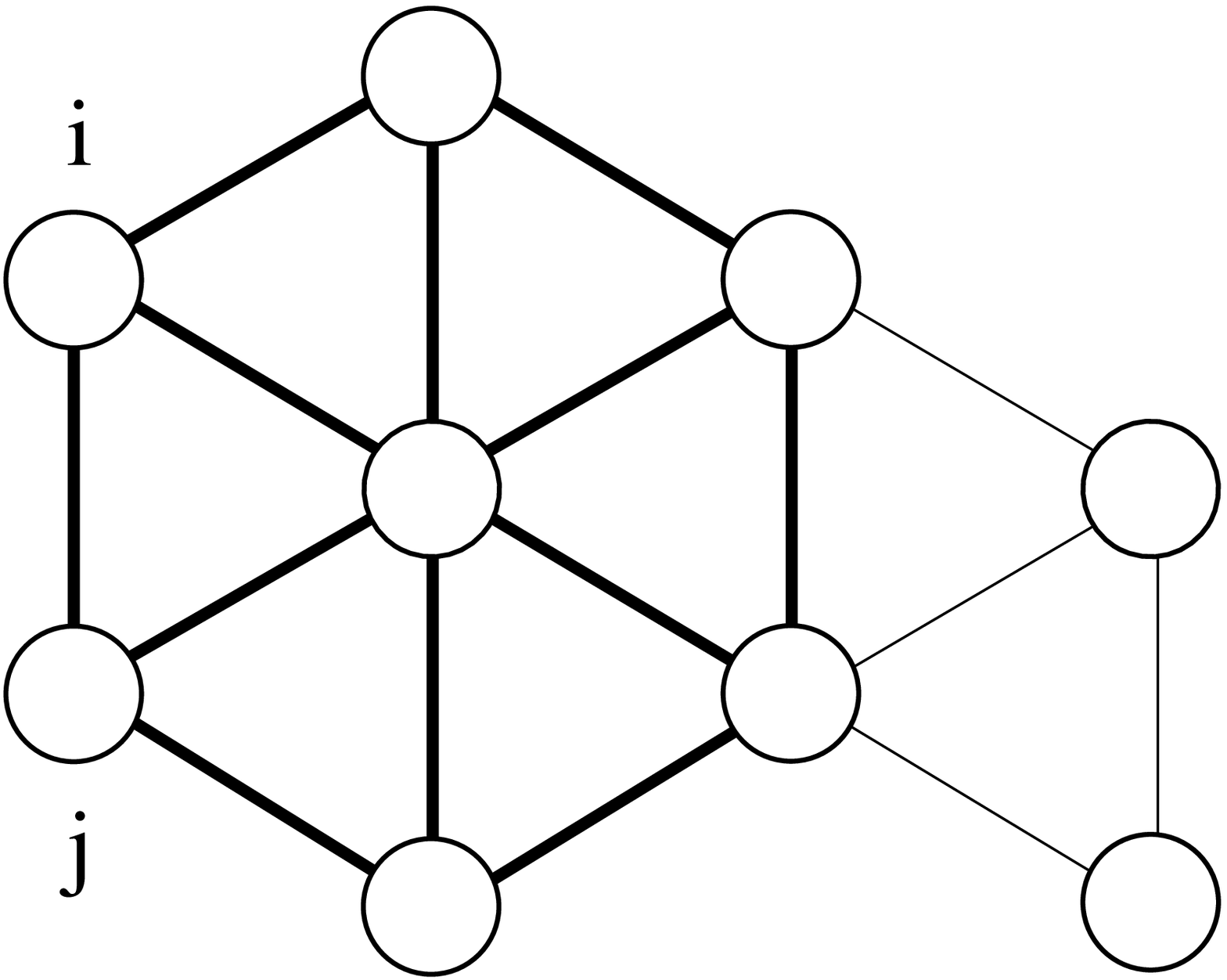,width=4.cm,angle=0}
}
\caption{{}In this two dimensional example, each node (circle) represents a
  point that has two positional degrees of freedom, while each bond fixes the
  distance between two points and thus provides one constraint. The graph on
  the left still has three remaining degrees of freedom: two translations and
  one rotation. Bond $ij$ (right) is a \emph{redundant} bond, because it
  connects two nodes which were already rigidly connected. After adding bond
  $ij$, the bonds that provided the rigid connection between $i$ and $j$
  become \emph{overconstrained} (thick lines). Any one of these can be removed
  without altering the number of remaining degrees of freedom, which still
  equals three. The graph on the right has one redundancy, but 12
  overconstrained bonds.}
\label{fig:redundancies}
\end{figure}

We consider a graph made of $N$ sites, each with $g$ degrees of freedom, and
for the moment assume that \hbox{$h=0$}, i.e.\ there are no ghost-field
constraints.  Each present bond removes one degree of freedom from the system,
unless it is a \emph{redundant} bond. A redundant bond (or constraint) is one
that links two nodes which were \emph{already} rigidly connected, as for
example nodes $i$ and $j$ in \Fig{fig:redundancies}.  The addition of a
redundant constraint does not reduce the number of degrees of freedom in the
system.  Thus the balance of degrees of freedom reads
\begin{equation}
N_F = N g - E + B_r,
\label{eq:fm}
\end{equation}
where $N_F$ is the number of uncanceled degrees of freedom~\footnote{Or zero
  modes, or \emph{floppy} modes.}, $E$ is the number of bonds in the graph,
and $B_r$ is the number of \emph{redundant} bonds.
\\
For an \ER\ graph one has \hbox{$\langle E \rangle =\coordn N/2$}.  Defining
\hbox{$\langle B_r\rangle = \coordn_r N/2$}, the average density
$\udof(\coordn)$ of zero modes per site is written
\begin{equation}
\udof(\coordn) = \frac{\langle N_F\rangle }{N} = g - \frac{1}{2} \left [
  \coordn   -\coordn_r(\coordn) \right ]. 
\label{eq:f}
\end{equation}
Let us assume that $b$ is a redundant bond. This means that its two end-nodes
are rigidly connected even if $b$ is removed. Let ${\Es}_b$ be the subset of
bonds that, in the absence of $b$, provide rigidity to its two end-nodes.
After adding $b$, any of the bonds in $\{\Es_b + b\}$ (thick lines in
\Fig{fig:redundancies}) can be removed without altering the total number of
degrees of freedom $N_F$.  The subset of bonds $\{\Es_b + b\}$ is said to be
\emph{overconstrained}. From a physical point of view, the overconstrained
bonds may be defined as those that carry an internal stress because of the
addition of a redundant bond $b$ that has a length-mismatch.
\\
Please notice the important difference between the number of redundancies and
the number of overconstrained bonds: when adding a redundant bond (like $ij$
in \Fig{fig:redundancies}) the number of redundancies $B_r$ always increases
by exactly one. However the number of overconstrained bonds may increase by
more than one. In the example shown in \Fig{fig:redundancies}, the number of
overconstrained bonds increases by twelve.
\\
We then conclude that, when adding \emph{any} bond to a graph, the number of
redundancies $B_r$ will either increase by one (if the chosen sites were
rigidly connected) or stay unchanged (if not), in which case the number of
zero modes $N_F$ will either stay unchanged or decrease by one.  This implies
that the densities $\coordn_r$ of redundant bonds and $\udof$ of zero modes
must be continuous functions of the density $\coordn$ of present bonds.  The
density of overconstrained bonds, on the other hand, needs not be continuous.
\\
Let us now discuss how the balance of degrees of freedom is modified by
ghost-field bonds (\Sec{sec:ghostfield}).  If a site has $n< g$ present ghost
edges connecting it to the rigid background, then it contributes with
\hbox{$(g-n)$} degrees of freedom to \Eqn{eq:fm}. If on the other hand it is
connected to the background by $n \geq g$ ghost bonds, then this site has no
degrees of freedom left to contribute to \Eqn{eq:fm}.  The effective number
$\geff(h)$ of remaining degrees of freedom contributed by an average site is
thus
\begin{eqnarray}
\geff(h) &=&  \sum_{n=0}^g (g-n) P_n(h)
\nonumber \\ 
&=& g (1-G_{g+1}(h)) - h (1-G_{g}(h)),
\label{eq:geff}
\end{eqnarray}
where $G_{g}$ is defined by (\ref{eq:F}). Taking this result into account, in
the presence of a field $h$ we now have that
\begin{equation}
\udof(\coordn,h) = \geff(h) - \frac{1}{2} \left [ \coordn-\coordn_r(\coordn,h)
\right ].
\label{eq:f2}
\end{equation}
Notice that $\udof$ and $\coordn_r$ are trivially related through the addition
of a known function of $\coordn$ and $h$. Therefore the knowledge of either
$\udof$ or $\coordn_r$ provides exactly the same physical information about
the system.
\\
The importance of (\ref{eq:f2}) is twofold: in the first place, as discussed
above, $n_F$ and $\coordn_r$ must be continuous functions of $\coordn$. A
similar reasoning allows one to conclude that they must also be continuous
functions of $h$. The continuity of $\coordn_r$ and $\udof$ is a key property,
and will be used later in \Sec{sec:locat-trans-point} to select the physically
correct solution of (\ref{eq:h2}) whenever indeterminacies arise.  Secondly,
in \Sec{sec:udof-free-energy} it will be argued that $n_F$ can be identified
with the logarithm of the partition function for the RP problem, and the
consequences of such identification will be discussed. Although this
identification is not necessary for solving the RP problem, it provides
interesting additional insight into this problem insofar it helps making a
link with thermodynamics.
\subsection{Calculation of $\coordn_r$}
\label{sec:calculation-coordn_r}
We now show how the density $\coordn_r(\coordn,h)$ of redundant bonds is
calculated. Obviously when $\coordn=0$ there are no redundant bonds, i.e.\ 
$\coordn_r(\coordn=0,h)=0$, so we can write
\begin{equation}
\coordn_r(\coordn,h) =  \int_{0}^\coordn \left .
\deriv{\coordn_r}{\coordn}  \right |_h
d\coordn ,
\label{eq:int}
\end{equation}
where the integral is done along a path of constant $h$.  In
\Apd{sec:field-p-derivatives} we show~\footnote{For alternative derivations
  see~\cite{JTG96,DJTF99}} that
\begin{equation}
\left .
\coordn \deriv{\langle B_r\rangle}{\coordn}  
\right |_{h}
= \langle
B_{ov}\rangle, 
\label{eq:deriv}
\end{equation}
where $\langle B_{ov}\rangle _\coordn$ is the total number of
\emph{overconstrained} bonds in $\Es$.  Consider two randomly chosen sites $i$
and $j$. As discussed in \Sec{sec:gammac}, a bond $b_{ij}$ is overconstrained
if $i$ and $j$ are rigidly connected to the background, even in the absence of
this bond. On a random graph, this happens with probability $R^2$.  Thus a
bond is present and overconstrained with probability $(\coordn/N)R^2$, and
therefore the average number of overconstrained bonds is $\langle
B_{ov}\rangle = \coordn R^2 N/2$.  We can now write \Eqn{eq:deriv} as
\begin{equation}
\left .
\deriv{\coordn_r}{\coordn}  \right |_{h} = R^2,
\label{eq:deriv2}
\end{equation}
so that now \Eqn{eq:int} reads
\begin{equation}
\coordn_r(\coordn,h) =  \int_{0}^{\coordn} 
R^2(\coordn,h) d\coordn.
\label{eq:int2}
\end{equation}
Notice that, while (\ref{eq:int}) and (\ref{eq:deriv}) are exact for arbitrary
graphs, (\ref{eq:deriv2}) only holds under the assumption that two neighboring
sites $i$ and $j$ are independently rigid with probability $R$.  Thus in
deriving (\ref{eq:int2}), correlations have been ignored. This is correct on
random graphs of the type considered here, but fails on finite dimensional
lattices.
\\
It is not immediately obvious how (\ref{eq:int2}) can be integrated since we
do not have an explicit expression for $R$ as a function of $\coordn$. However
using $x=\coordn R=y-h$ one can write $R^2 d\coordn = R dx -x dR = (R- (y-h)
\partial R/\partial y) dy$, where $R=G_g(y)$ as given by (\ref{eq:h2}). After
integrating by parts,
\begin{eqnarray}
\coordn_r &=& \left . \left \{ 2 \int G_{g}(y) dy -  (y-h)G_{g}(y)  \right \}
\right |_{y=h}^{y=x+h}
\label{eq:intec}
\end{eqnarray}
Using (\ref{eq:F}), \Eqn{eq:intec} can be integrated to give
\begin{eqnarray}
\coordn_r 
&=& \left . 
\left \{(y+h-2g)G_g(y) + 2gP_g(y)  \right \}
\right |_{y=h}^{y=x+h}
\nonumber \\ 
&=& (y+h) G_{g}(y) -  2h G_{g}(h) + 2 g \left [ G_{g+1}(h) - G_{g+1}(y) \right ] 
\nonumber \\ 
\label{eq:gammarx}
\end{eqnarray}
This expression enables us to calculate the density of redundant bonds
analytically. In previous work on Bethe lattices~\cite{DJTF99},
(\ref{eq:int2}) was solved in zero field by numerical integration, using for
this purpose the values of $R$ that result from iteration of (\ref{eq:bd3}).
Although the algebra becomes slightly more complicate, the approach used here
to calculate $\coordn_r$ is applicable to Bethe lattices as well. Later in
\Sec{sec:case-zero-field} we present some results for Bethe lattices.
\subsubsection{The density of zero modes}
Using (\ref{eq:h2}), (\ref{eq:geff}), (\ref{eq:f2}), and (\ref{eq:gammarx}) we
can write
\begin{equation}
\udof(\coordn,h) = \frac{\coordn}{2} (R^2-1) + (h-g) (R-1) + g P_g(y)
\label{eq:udoff}
\end{equation}
where $y=x+h=\coordn R +h=\coordn (R+H)$. Because of (\ref{eq:f2}), continuity
of $\coordn_r$ implies that of $\udof$.  
\subsubsection{A consistency check}
In the limit $\coordn \to N$, all bonds are present and there are no remaining
degrees of freedom, i.e.\ $\udof \to 0$.  It is easy to verify that
\Eqn{eq:udoff} indeed satisfies this condition.
\begin{figure}[h]
  \centerline{\psfig{figure=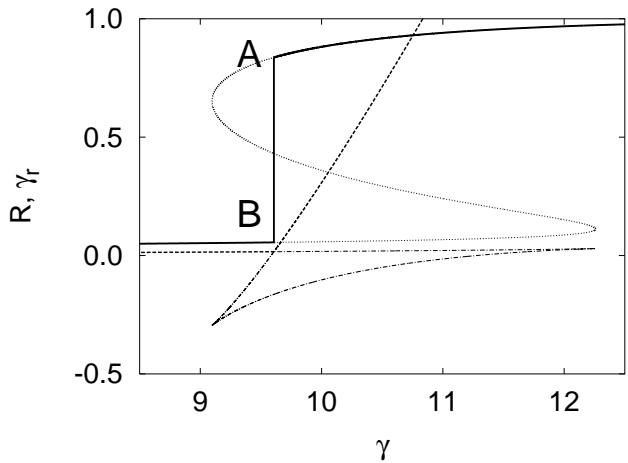,width=9cm,angle=270} }
  \centerline{}\caption{{}The density of redundant bonds $\coordn_r$ (thick
    dashed) describes a Maxwell loop (dot-dashed) whose crossing point defines
    the location of the true transition. After discarding the unphysical loop,
    the density $R$ of rigid sites (thick solid) jumps at the transition
    between A and B, but $\coordn_r$ is continuous.  Shown is an example with
    $g=10$ and $h=5$.}
\label{fig:jump}
\end{figure}
\section{Location of the transition point}
\label{sec:locat-trans-point}
\subsection{The general case}
\label{sec:general-case}
\begin{figure}[h]
  \centerline{\textbf{a)}\psfig{figure=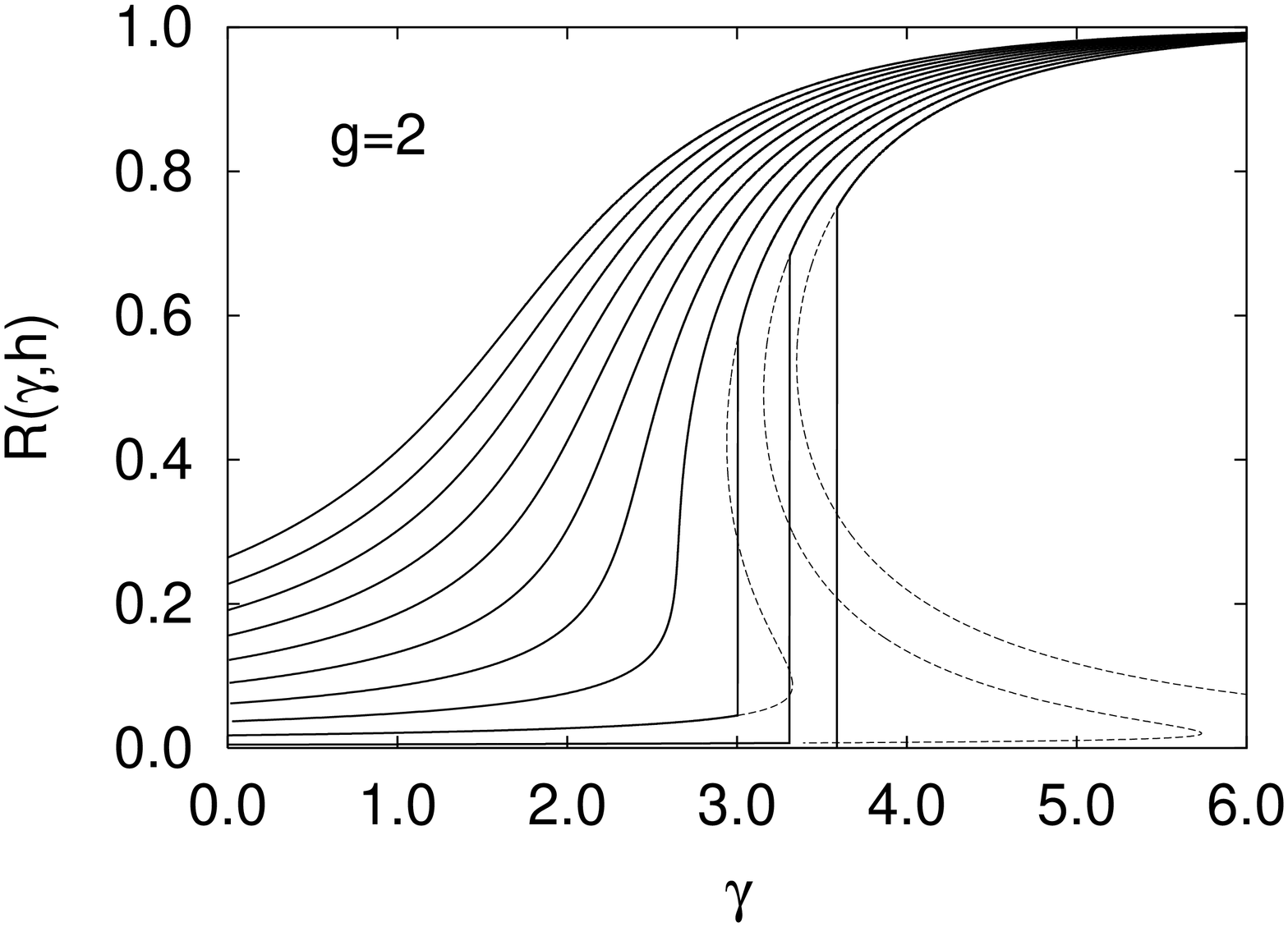,width=9cm,angle=270}}
  \centerline{\textbf{b)}\psfig{figure=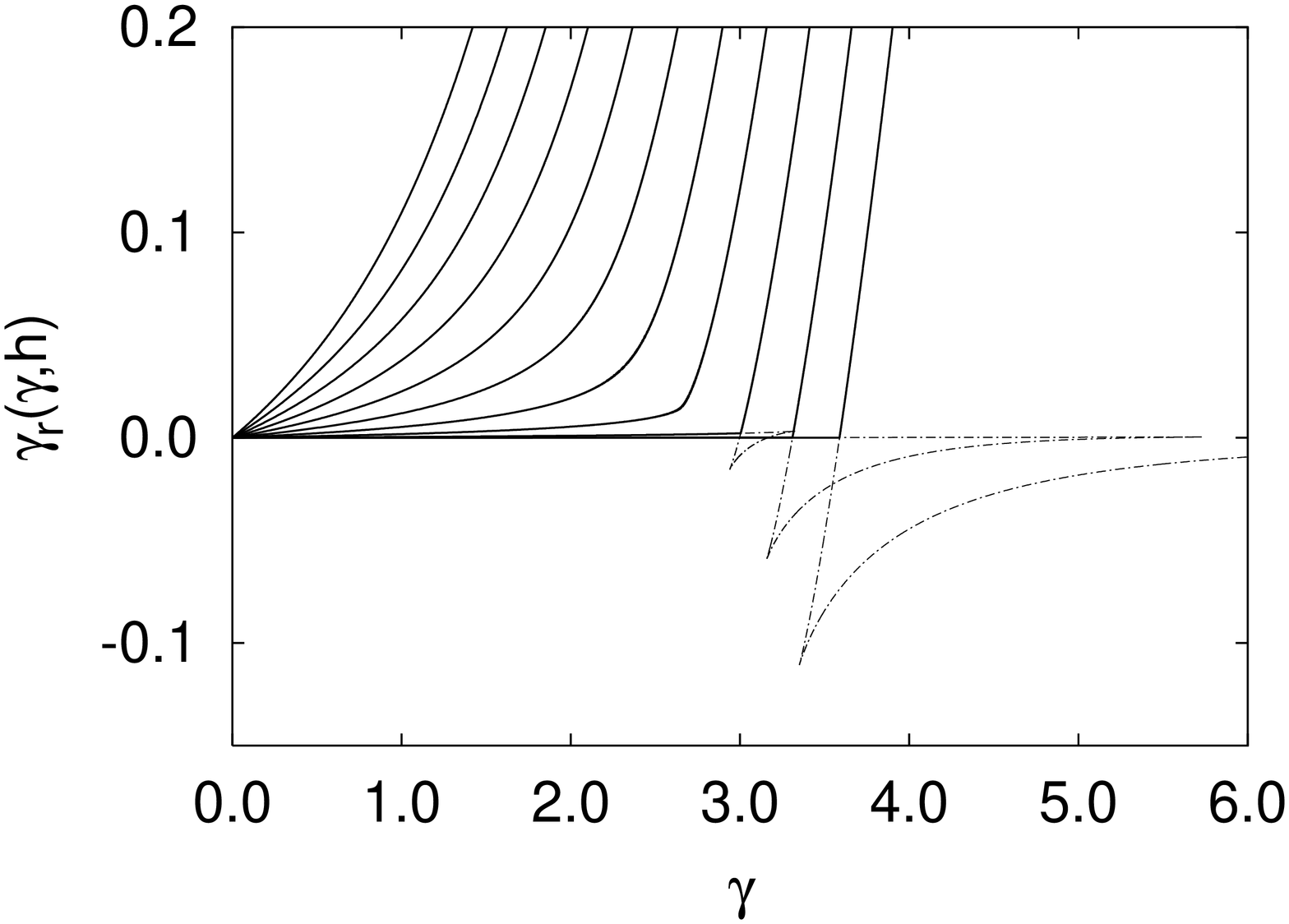,width=9cm,angle=270}}  
\caption{{} \textbf{a)} The final values for the density of
  rigid sites $R$ (thick lines). The transition point $\coordn_c(h)$, where
  $R$ undergoes a discontinuous jump, is determined by the physical
  requirement that the density of redundant bonds $\coordn_r$ (\textbf{b)}) be
  continuous.  In this example $g=2$ and (from right to left):
  $h=0.0,0.1,0.20,\ldots,1.0$.  For $g=2$ the critical field is $h_c=0.287$.
  }
\label{fig:tcorr}
\end{figure}

The value $\coordn_c(h)$ where the transition takes place is univocally
determined by the requirement that $\coordn_r$ be a continuous function of
$\coordn$. In order to see how this works, imagine that for a given $h<h*$ one
lets $h<y<\infty$ and calculates $\coordn_r(y,h)$ (\Eqn{eq:gammarx}) and
$\coordn(y,h)$ (\Eqn{eq:bd5}), thus obtaining a plot of $\coordn_r$ vs
$\coordn$ (\Fig{fig:jump}). In doing so one finds that $\coordn_r$ describes a
loop (sometimes called a ``Maxwell loop'' in Statistical Mechanics).
\\
The existence of a crossing point implies that two values $y_A$ and $y_B$
exist such that
\begin{equation} 
\left . \left \{
(y+h-2g) G_{g}(y) +  2 g P_{g}(y)
\right \} \right |^{y=y_B}_{y=y_A} = 0,
\label{eq:crossing2}
\end{equation}
where $y_{A,B}=\coordn_c(h) R_{A,B}+h$.
\begin{figure}[h]
\centerline{\psfig{figure=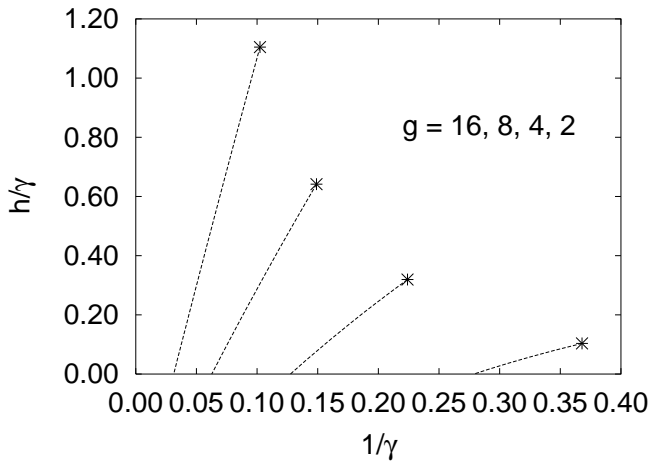,width=9cm,angle=270}}
\caption{{} Coexistence lines (lines of first-order phase transitions--dashed)
  in the space defined by the two intensive parameters of Rigidity
  Percolation: the ``field'' $H=h/\coordn$ and the ``temperature''
  $1/\coordn$, for several values of the number $g$ of degrees of freedom per
  site. The critical point, defined by \protect (\ref{eq:critpoint}), is
  indicated with an asterisk. For all $g\geq 2$, i.e.\ for Rigidity
  Percolation, this critical point has classical indices $\alpha=0$,
  $\beta=1/2$, $\gamma=1$ and $\delta=3$.  The case $g=1$ (Scalar Percolation,
  not shown) has its critical point at $\coordn=1$ and $H=0$, and belongs to a
  \emph{different} universality class, with $\alpha=-1$, $\beta=1$, $\gamma=1$
  and $\delta=2$.}
\label{fig:hcline}
\end{figure}
\Eqn{eq:crossing2} cannot be solved analytically in general, however it is
easily solved numerically (by iteration) for each value of $h$ and $g$, to
obtain $\coordn_c(g,h)$. In this fashion the location of the coexistence line
can be calculated to arbitrary precision (\Fig{fig:hcline}).  At the
transition point $\coordn_c(h)$ the order parameter $R$ jumps
(\Fig{fig:tcorr}a), but the density of redundant bonds $\coordn_r$ is
continuous (\Fig{fig:tcorr}b).
\subsection{The case of zero field. Maxwell counting on the
  $(g+1)$-rigid-core}
\label{sec:case-zero-field}
For zero field one has that $R=0$ for $\coordn<\coordn_c$.  \Eqn{eq:int2} then
implies that $\coordn_r=0$ for $\coordn \leq \coordn_c$. At the transition
point $\coordn_c$, a rigid cluster suddenly appears that contains a fraction
$R_c$ of the graph.  In zero field, \Eqn{eq:crossing2} reads
\begin{equation}
(x_c - 2g )R_c + 2 g P_g(x_c) = 0,
\label{eq:crossing3}
\end{equation}
where all quantities are evaluated in the rigid phase ($R$ and $y$ are zero in
the floppy phase).
\\
Even this simpler equation does not admit analytic treatment. However, very
precise solutions can be obtained numerically as described already.
\Tbl{tbl:gammac} contains some examples.
\\
\begin{table}
\begin{center}
\begin{tabular}{ccc}
\toprule
$g$ & $\coordn_c(g)$ & discontinuity $R_c$ \\
\colrule
$2$ & $3.58804747296539\ldots$ & $0.74915378510250\ldots$ \\ 
\colrule
$3$ & $5.7549256115462\ldots$ & $0.8812398370507\ldots$ \\
\colrule
$4$ & $7.84295819428849\ldots$ & $0.933538491167168\ldots$ \\
\colrule
$5$ & $9.89551361859910\ldots$ & $0.959638549916489\ldots$ \\
\colrule
$6$ & $11.9288724790262765\ldots$ & $0.974275720961391\ldots$ \\
\colrule
$\to \infty$ & $2g$ & $1.0$ \\
\botrule
\end{tabular}
\end{center}
\caption{{}The Rigidity Percolation threshold coordination $\coordn_c$, and the
  jump in the rigid cluster density $R_c$ on \ER\ random graphs, obtained from
  solving \protect   (\ref{eq:crossing3}) for several values of $g$, the
  number of degrees of freedom per site.}
\label{tbl:gammac}
\end{table}
Please notice that $\coordn_c$ is \emph{smaller} than the value $2g$ needed
for global balance of constraints and degrees of freedom. However, all degrees
of freedom belonging to sites \emph{on the rigid cluster} (which appears
suddenly at the transition point) are by definition canceled out by
constraints.  Moreover, since for zero field the number of redundant
constraints per site is exactly zero at the transition (See \Fig{fig:tcorr}),
one concludes that the average number of bonds per site, for sites \emph{on
  the rigid cluster}, must be exactly $2g$ at $\coordn_c$. The conclusion is
then that sites on the rigid cluster have more bonds than average.
\\
This observation may be formalized in the following way.  Using $\coordn R=x$,
$R=\sum_{k=g}^{\infty} P_k(x)$ with $P_k(x)=e^{-x}x^k/k! $, and
$xP_k(x)=(k+1)P_{k+1}(x)$, condition (\ref{eq:crossing3}) can be written
\begin{equation}
\sum_{k=g+1}^{\infty} (2g-k) P_k(x)=0,
\label{eq:cc0}
\end{equation}
which in turn implies
\begin{equation}
\frac{\sum_{k=g+1}^{\infty}k P_k(x)}{\sum_{k=g+1}^{\infty}P_k(x)}=2g.
\label{eq:cc1}
\end{equation}
This sum counts the number of rigid neighbors for sites on the
$(g+1)$-rigid-core, that is, the subgraph of the rigid cluster that has minimum
coordination $(g+1)$.  \Eqn{eq:cc1} means that the $(g+1)$-rigid-core has an
exact balance of degrees of freedom.
\\
Clearly, the rigid cluster may also contain sites not in the
$(g+1)$-rigid-core. When these are considered, it is possible to see that the
balance of constraints and degrees of freedom is still respected. 
\subsubsection*{The case of large $g$}
An analytic solution of \Eqn{eq:crossing3} is possible (though somewhat
trivial) in the $g>>1$ limit.  In this case $P_g \to 0$ and $R_c \to 1$, i.e.
there is a jump from $R=0$ to $R=1$ at the transition point. Plugging these
observations into (\ref{eq:crossing3}) one gets, for zero field,
\begin{equation} 
\coordn_c = 2g \qquad g>>1.
\label{eq:glarge}
\end{equation}
This means that the rigid transition happens exactly at the point where global
balance between constraints and degrees of freedom is attained. In other
words, at $\coordn_c$ Maxwell counting is exact on the rigid cluster for all
$g$, while for $g>>1$ it is globally exact as well.
\subsubsection*{Bethe lattices}
A graph where each site has exactly $z$ randomly chosen neighbors is
topologically equivalent to a Bethe lattice, except for finite-size
corrections. When this system is randomly diluted, one obtains what is called
``random bond model'' in Ref.~\cite{DJTF99}, in which each of the $z$ bonds on
a site is present with probability $p$. We will only state the final results
for this system. In zero field the transition point $p_c$ is determined by
\begin{equation}
\sum_{j=g+1}^z (j-2g) Q_j^{(z)}(x_c)=0,
\label{eq:bethe0}
\end{equation}
where $x=pT$, $Q_j^{(z)}(x)={z \choose j} x^j(1-x)^{z-j}$ and $T$ satisfies
the recursive equation~\cite{MDLF97}
\begin{equation}
T=\sum_{j=g}^{z-1} P_j^{(z-1)}(x).
\end{equation}
The transition condition (\ref{eq:bethe0}) gives rise to a polynomial equation
which is easily solved for small values of $g$ and $z$. For $g=2$ and $z=5$
the corresponding quadratic equation results in $p_c=0.83484234$.  For $g=2$
and $z=6$ a cubic equation is obtained, and its solution is $p_c=0.656511134$.
This last value is consistent with but more precise than $p_c=0.656$ as
obtained in previous work~\cite{MDLF97} using numerically exact matching
algorithms for RP~\cite{MA96,JHA97}.
\section{Relation between $\udof$ and a free energy}
\label{sec:udof-free-energy}
In this Section, the consequences of making the identification \hbox{$\udof
  \to (\log \partfunc)/N$} are explored.  It has been shown some time ago that
SP can be mapped onto the $q\to 1$ limit of the Potts model\cite{KFP69,FKR72}.
This mapping allows one to draw a parallel between an equilibrium
thermodynamic transition (Potts) and percolation, which may be described as a
purely geometric transition. One of the outcomes of this equivalence is the
identification of the total number of clusters in percolation as the logarithm
of the partition function for the resulting Potts model. The existence of this
mapping has had profound impact on our understanding of scalar percolation,
e.g.\ by allowing the exact calculation of percolation critical indices in two
dimensions~\cite{BHF96}.  No equivalent mapping has been found for RP yet.
However it has been suggested~\cite{JTG96,DJTF99} that a possible
generalization for the free energy in RP for zero field is the total number of
remaining degrees of freedom (zero modes or \emph{floppy modes}). In the $g=1$
case of RP each cluster has one remaining degree of freedom, thus the SP
result is recovered.
\\
In \Sec{sec:gammac} it was shown that $\udof$ (equivalently $\coordn_r$) has
to be continuous. This was used in \Sec{sec:locat-trans-point} to locate the
point $\coordn_c$ where the first-order transition happens. Thus $\udof$ plays
a role similar to that of the Gibbs free energy, which is also a continuous
function of its intensive parameters.  In this section we add further evidence
supporting the identification of $\udof$ as the logarithm of the ``partition
function'', or free energy, for the RP problem on random graphs.  We
demonstrate that the \emph{ansatz} $\partfunc_{RP} \propto e^{N\udof}$ is
consistent also in the presence of a field, and show how several thermodynamic
quantities result from derivatives of this free energy.
\\
In \Apd{sec:derivatives-coordn_r} it is shown that the derivatives of $\udof$
are
\begin{subequations}\label{eq:dernf}
\begin{eqnarray}
\left . \fg \right |_h &=&  \frac{1}{2} (R^2 -1)
\\
\left . \fh \right |_\coordn &=&  R-1,
\label{eq:ordpar}
\end{eqnarray}
\end{subequations}
As expected, the order parameter $R$ results form deriving the logarithm of
the partition function with respect to the field-like parameter, up to a
constant shift.
\\
From \Eqns{eq:dernf} we furthermore obtain
\begin{subequations}
\begin{eqnarray}
\left . \fg \right |_H &=&  \frac{1}{2} (R^2 -1) + H (R-1)
\\
\left . \fH \right |_\coordn &=&  \coordn (R-1).
\end{eqnarray}
\end{subequations}
\subsection{Maxwell's Rule of Equal Areas}
\label{sec:maxwells-rule-equal}
\begin{figure}[h]
\centerline{\psfig{figure=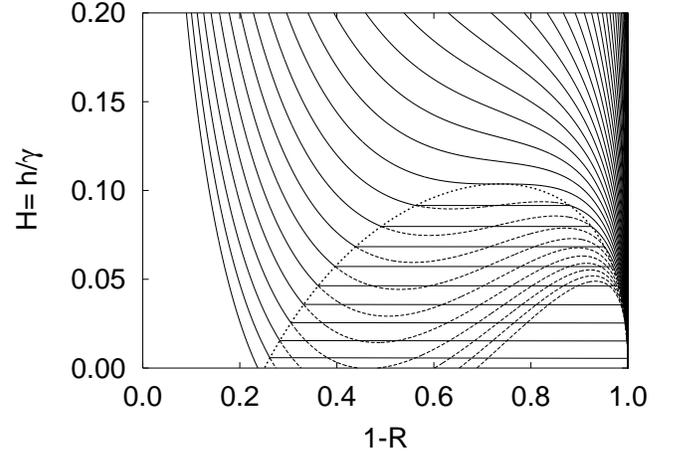,width=9cm,angle=270}}
\caption{{} This plot of the field $H$ versus the order parameter $R$ for
  Rigidity Percolation ($g=2$ in this example) is the equivalent of a $P$-$V$
  diagram for fluids. Shown are the ``isotherms'' (solid lines) along which
  the coordination number $\coordn$ is constant. The thick dashed line
  delimits the ``coexistence'' region, and was obtained by solving
  (\ref{eq:equalarea}) numerically. The areas delimited by the thin dashed
  lines above and below the (solid) horizontal coexistence line are equal.}
\label{fig:Hvsr.g.2}
\end{figure}

The continuity requirement (\ref{eq:crossing2}) employed in this work to locate
the true transition point $\coordn_c$ is analogous to the condition that the
free energy be continuous, as noted in previous work~\cite{DJTF99}. The
continuity of an appropriate thermodynamic potential is discussed in
elementary statistical mechanics textbooks in connection, for example, with
Van der Waals' equation~\cite{SI87,JNAT02}, where it is used as a criterion to
locate the transition.  The continuity condition for our RP problem is written
in general as
\begin{equation}
\int_{A}^{B} d \udof=0,
\label{eq:continuity}
\end{equation}
where $A$ and $B$ are the two phases that coexist at the first-order
transition (see \Fig{fig:jump}).  Clearly (\ref{eq:crossing2}) is the
particular case of (\ref{eq:continuity}) for which the integration is done
along a path of constant field.  Choosing a path $A \to B$ on which $\coordn$
is constant instead, it is easy to put (\ref{eq:continuity}) in the form of
the famous Maxwell's ``rule of equal areas''. Start from
\begin{equation}
0=\int_{A}^{B} \left .\fh \right |_{\coordn} dh,
\end{equation}
where the integral is done along an ``isotherm'' (a line of constant
$\coordn$).  After using (\ref{eq:ordpar}) and integrating by parts, this
reads
\begin{equation}
h(R_B-R_A)=\int_{A}^{B} h(R,\coordn) dR,
\label{eq:equalarea}
\end{equation}
where $h(R,\coordn) = G_{g}^{-1}(R) -\coordn R$.
\\
\Eqn{eq:equalarea} is equivalent to the condition $P(V_B-V_A)=\int_{A}^{B}
P(V,T) dV$ for a fluid, i.e.\ Maxwell's rule.  \Fig{fig:Hvsr.g.2} serves to
illustrate the fact that the areas above and below the horizontal coexistence
line are equal, as (\ref{eq:equalarea}) requires.
\\
\subsection{Energy, Entropy, and Work}
In the following we will assume that $\udof$ (\Eqn{eq:udoff}) is the logarithm
of the partition function, i.e.\ $\udof = -\coordn f$ with $f$ a free-energy
density, and $\coordn =1/T$ the inverse temperature. The energy density $e$ is
then
\begin{equation}
e  = \left . -\frac{\partial \udof }{\partial \coordn } \right |_H = 
\frac{1}{2} (1-R^2) + H (1-R).
\label{eq:e}
\end{equation}
The entropy per site $s$ turns out to be
\begin{equation}
s = - \left . \deriv{f}{T} \right |_H = 
\udof + \coordn e =
g \left [ P_g(y) -(R-1)\right ].
\label{eq:s}
\end{equation}
From the first law of thermodynamics, and using \Eqns{eq:e} and (\ref{eq:s}),
we conclude that, in an infinitesimal transformation, the ``work'' done by the
system against the environment is:
\begin{equation}
dw = dq -de = \frac{1}{\coordn} ds -de = \frac{1}{\coordn} \deriv{\udof}{H} dH
= (R-1) dH
\label{eq:w}
\end{equation}
The analogous of the constant-pressure heat capacity for a fluid is in our
case the constant-field heat capacity
\begin{equation}
c_H = \left . -\coordn \deriv{s}{\coordn} \right |_H 
= \left . -\coordn^2 \deriv{e}{\coordn} \right |_H =
\frac{y^2 P_{g-1}(y)}{1-\coordn P_{g-1}(y)},
\end{equation}
which is nonnegative by (\ref{eq:stability2}), and diverges as $|\coordn -
\coordn^*|^{-1}$ at the critical point.  This of course should not be taken to
mean that $\alpha=1$, since it is the constant-density heat capacity what
defines $\alpha$ (For a discussion see Ref.~\cite{LBS99}).  The constant-field
heat capacity $c_H$ is expected to diverge with the same exponent $\gamma$ as
the susceptibility, unless the system has certain symmetries, which is the
case for the Ising model, but not for RP.
\\
The analogous of the ``constant-density'' heat capacity is
\begin{equation}
c_R = \left . -\coordn \deriv{s}{\coordn} \right |_R,
\label{eq:Cv}
\end{equation}
which is zero and thus $\alpha=0$.
\subsection{Clausius-Clapeyron Equation}
Let $\Delta f = -1/\coordn \Delta \udof = -1/\coordn (\udof(B) - \udof(A)) $
be the free-energy difference between the two phases in the coexistence
region. This is a function of the ``temperature'' $T=1/\coordn$ and the field
$H=h/\coordn$. Following standard texts~\cite{HS87}, we write
\begin{equation}
\left . \deriv{H}{T} \right |_{\Delta f} = 
- \frac{ \left . \deriv{\Delta f}{T} \right |_H}
{\left . \deriv{\Delta f}{H} \right |_T}
= - \frac{\Delta s}{\Delta R} ,
\end{equation}
where the derivative on the left-hand side is evaluated on any line of
constant $\Delta f$. Using (\ref{eq:s}) one can write $\Delta s=g(\Delta P_g -
\Delta R)$. On the coexistence line, $\Delta f=0$ so $\left . \Delta s \right
|_{coex}= \left . \coordn \Delta e \right |_{coex}$ and, using (\ref{eq:e}) we
have that $\left . \Delta s \right |_{coex} = - h \Delta R -\coordn/2
\Delta(R^2) =-\Delta R (\frac{y_A+y_B}{2})$. Thus the Clausius-Clapeyron
equation for the RP problem can be written as
\begin{equation}
\left . \deriv{H}{T} \right |_{coex} = 
 \frac{ y_A + y_B}{2}
\end{equation}
Since on approach to the critical point, $y\to y^*=g-1$ (\Eqn{eq:critpoint}),
we conclude that the coexistence line reaches the critical point with slope
$(g-1)$ in the $H$-$T$ plane (See~\Fig{fig:hcline}).
\section{Discussion}
\label{sec:discussion}
The Rigidity Percolation problem with $g$ degrees of freedom per site has been
considered on \ER\ graphs with average coordination $\coordn$. An external
field $h$ is introduced by connecting each site to a rigid background, or
``ghost site'', with a certain probability that depends on $h$ as described in
\Sec{sec:ghostfield}.  The resulting Equation of State (\ref{eq:h2}) for the
density $R$ of sites that are rigidly connected to the background, undergoes a
first-order phase transition on a ``coexistence line'' $\coordn_c(h)$. This
line ends at a critical point (\Fig{fig:hcline}) with classical exponents
$\alpha=0$, $\beta=1/2$, $\gamma=1$ and $\delta=3$. For comparison, the
critical point of Scalar percolation (the case of $g=1$) is located at $h=0$
and has different critical exponents: $\alpha=-1$, $\beta=1$, $\gamma=1$ and
$\delta=2$~\cite{SAI94}. Therefore in the MF approximation, Scalar and
Rigidity percolation are in different universality classes. In two dimensions,
a similar conclusion is reached by numerical means~\cite{JTG96,MDC99}, were
both transitions are continuous in zero field.
\\
It has been recently argued~\cite{BCR02} that for certain spin systems, the
existence of a discontinuous transition in the MF approximation is enough to
ensure that, in finite space dimensions $d$, there is a discontinuous
transition if $d$ is large enough. If a this result holds for RP, one would
expect to see a discontinuous transition in zero field for $d$ (and perhaps
$g$) large enough.  Up to now, extensive numerical simulations for RP have
been performed only in two dimensions~\cite{MDC99,MDS95,JTG96,JTG95}. Except
for pathological cases where the transition happens at zero
dilution~\cite{MDC99}, it seems that the transition is continuous for all $g$
in two dimensions. Large scale simulations are still needed to clarify this
issue.
\\
On Cayley Trees with a boundary, $g$-rigidity percolation(RP) is closely
related to $(g+1)$-bootstrap percolation(BP)~\cite{CLRB79,MDLF97}. Both
percolate at the same density $p$ of present bonds, although with different
spanning cluster densities on trees.  On random graphs, which have the local
structure of a tree but no boundaries, $k$-core transition stays unchanged,
but the RP transition is delayed to larger $p$ values~\cite{MDLF97}. We have
explicitely shown that, at the critical concentration $\coordn_c$ where the
rigid cluster first appears, it has an exact balance of degrees of freedom,
i.e.\ it has exactly $g$ bonds per site. This condition is satisfied both on
the whole rigid cluster and on its $(g+1)$-rigid core. However globally the
system has lesser bonds than needed to attain this balance, meaning that the
rigid cluster is selectively made out of sites with more bonds than average.
\\
In previous work on Bethe lattices in zero field~\cite{DJTF99} it was
suggested that the number $\udof$ of uncanceled degrees of freedom is a free
energy for the RP problem. For SP ($g=1$) this result holds exactly, being one
of the outcomes of the Fortuyn-Kasteleyn Random-Cluster
model~\cite{KFP69,FKR72}, which contains SP and the Potts model as particular
cases. For RP, however, this identification only has the status of a plausible
\emph{ansatz}.  This \emph{ansatz} was used recently to predict some
thermodynamic properties of chalcogenide glasses~\cite{NC00}, for which the RP
transition has been shown to be relevant~\cite{PT79,PTC85,TR85,TDR99}.
\\
Some of the reasons to believe that this identification might be correct in
general are: a) the fact that $\udof$ must be continuous at the discontinuous
RP transition, and, b) the fact that for $g=1$, $\udof$ is the number of
connected clusters per site, so the FK result for SP is recovered exactly. In
this work we have shown that, in the presence of a properly defined field $h$,
the ordered parameter $R$ can be obtained as a derivative of the free energy
with respect to $h$, thus adding further support to the belief that this
identification is correct.
\\
Under the assumption that $\log \partfunc = N \udof$, the entropy per site $s$
can be derived. The resulting expression (\ref{eq:s}) was found to depend on
the order parameter $R$ alone. This is a property of other MF systems like for
example the Ising ferromagnet.
\\
\begin{figure}[h]
\centerline{\psfig{figure=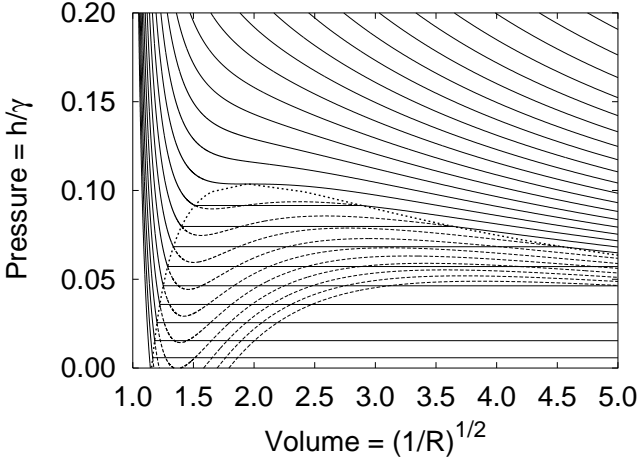,width=9cm,angle=270}}
\caption{{} Illustration of the similarities between RP and a
  condensation transition. In this plot, $H=h/\coordn$ is taken to be the
  analog of a pressure and $1/R^{1/2}$ the analog of a volume. The data are
  the same as shown in \protect \Fig{fig:Hvsr.g.2}.  }
\label{fig:Hvsrinv}
\end{figure}
On the pedagogical side, we have attempted to situate the discussion in terms
of the parallel between the RP transition in a field and a condensation
transition. Since only topological (connectivity) properties are important for
RP, it can be said that, in a sense, RP in a field is a sort of ``geometric
condensation transition''.  \Fig{fig:Hvsr.g.2} illustrates the similarities
between both transitions. However a closer match is possible. Comparing
(\ref{eq:h2}) to (\ref{eq:vdwmf}) one notices that, in a MF condensation
transition, $\lambda \propto \rho^2=1/v^2$ plays the role of the order
parameter $R$ in RP.  From this point of view, the analog of a ``volume'' in
RP would be $1/R^{1/2}$.  The analog of a $P$-$V$ plot is shown in
\Fig{fig:Hvsrinv}.  Notice however that, within this definition of a
``volume'', $P(V)$ does \emph{not} satisfy Maxwell's rule of equal areas,
$\int_A^B V(P) dP=0$.  
\acknowledgements The author has benefited from many useful discussions with
Prof.~Phillip Duxbury in the early stages of this work.  Financial support
from the SNI program of CONACYT, and from a CONACYT grant for research project
36256-E, is acknowledged.
\appendix
\def \func {{\mathcal{F}}}
\section{Derivatives of the number of redundant bonds}
\label{sec:field-p-derivatives}
Let $\func(\Es)$ be an arbitrary function of the set of present edges $\Es$ of
a graph $\graph$, and let $P(\Es)$ be the probability to have edge set $\Es$.
The average of $\func$ is defined as
\begin{equation}
\langle \func \rangle = \sum_{\left\{\Es \right\}} P(\Es) \func(\Es),
\label{eq:favg}
\end{equation}
where $\sum_{\left\{\Es \right\}}$ is a sum over all configurations of present
edges.
\\
If each edge is independently present with probability $p$ and absent with
probability $1-p$, then clearly
\begin{equation}
P_p(\Es) = p^{|\Es|} (1-p)^{|\Es_{max}|-|\Es|},
\label{eq:Pp}
\end{equation}
where $\Es_{max}$ is the edge set of maximum possible cardinality.
\\
We now write $p= p_1 p_2$ with $0\leq p_1, p_2 \leq 1$. This can be realized
by assuming that two types of bonds are independently present with
probabilities $p_1$ and $p_2$ respectively, and that in order for an edge to
be ``active'', both types of bonds must be present on that edge. $\func$ is
now a function of active bonds only, which are present with probability
$p_1p_2$, so that one can write
\begin{eqnarray}
\langle \func \rangle_{p_1p_2} &=& 
\sum_{\left\{\Es \right\}} P_{p_1}(\Es)
\sum_{\left\{\Es' \right\} \subseteq \Es} P_{p_2}(\Es')
\func(\Es')\nonumber \\ 
&=& \sum_{\left\{\Es \right\}} P_{p_1}(\Es) S(p_2,\Es)
\label{eq:fav}
\end{eqnarray}
Notice that $P_{p_2}(\Es')$ has a factor $p_2$ for each present bond in
configuration $\Es'$, and a factor $1-p_2=q_2$ for each bond in $\Es$ which is
absent in $\Es'$.  Factors associated with bonds not in $\Es$ can be summed
out to give one and therefore need not be considered. Thus if $|\Es|$ is the
number of present bonds in configuration $\Es$ and $|\Es'|$ the number in
$\Es'$, $P_{p_2}(\Es') = p_2^{|\Es'|}q_2^{|\Es|-|\Es'|}$.
\\
Now let $p_2 \to 1$. In this case we can expand $S(p_2,\Es)$ in powers of
$q_2$.  The zeroth-order contribution comes from $\Es'\equiv \Es$, the
first-order contribution from the $|\Es|$ configurations $\Es'$ which have
exactly one bond $b$ less than $\Es$, etc.
\begin{eqnarray}
S(p_2,\Es)&=& p_2^{|\Es|} \func(\Es) 
+ q_2 p_2^{|\Es|-1}  \sum_{b \in \Es}  \func(\Es-b)  + \nonumber \\
&+& q_2^2 p_2^{|\Es|-2}\sum_{b,b' \in \Es}   \func(\Es-b-b') + \ldots ,
\label{eq:S1}
\end{eqnarray}
where we have written $\Es -b$ to denote the result of deleting bond $b$ from
$\Es$. To first order in $q_2$,
\begin{equation}
S(p_2,\Es)=  \func(\Es) + q_2 \sum_{b \in \Es} \left( \func(\Es-b) - \func(\Es) \right),
\label{eq:S2}
\end{equation}
therefore
\begin{equation}
\langle \func \rangle_{p_1p_2} = \langle \func \rangle_{p_1} + 
q_2 \left\langle 
\sum_{b \in \Es} \left( \func(\Es-b) - \func(\Es) \right)
\right\rangle_{p_1} .
\label{eq:final}
\end{equation}
Taking derivatives with respect to $p_2$ and letting $p_2=1$ one finally finds
that, for arbitrary $\func$,
\begin{equation}
p \deriv{\langle \func \rangle}{p} = 
\left\langle 
\sum_{b \in \Es} 
\left \{
\func(\Es)-\func(\Es -b)
\right \} 
\right\rangle .
\label{eq:derp}
\end{equation}
\Eqn{eq:derp} generalizes previous results of Coniglio in Ref.~\cite{CC82},
and is the tool we use in this Section to derive some important relations.
\\
Consider now the case in which $\func=B_r(\Es)$, the total number of redundant
constraints. In order to calculate the derivative of $B_r$ with respect to
$p$, we notice that when removing a bond $b$ from a configuration $\Es$ of
present bonds, the total number of redundant constraints will be reduced by
one if and only if $b$ is \emph{overconstrained}. Otherwise if $b$ is not
overconstrained, $B_r$ remains unchanged by the removal of $b$. \Eqn{eq:derp}
then implies that
\begin{equation}
p \deriv{\langle B_r \rangle}{p} = 
\coordn \deriv{\langle B_r \rangle}{\coordn} = 
\langle B_{ov}\rangle , 
\label{eq:derbr}
\end{equation}
where $\langle B_{ov}\rangle $ is the average number of \emph{overconstrained}
bonds in $\Es$.
\\
A calculation of the field derivative of $\langle B_r \rangle $ requires some
additional considerations.  \Eqn{eq:derp} was derived under the assumption
that each edge is independently present with some fixed probability, while our
definition of ghost-field in \Sec{sec:ghostfield} implies that each site can
have \emph{any} number $n$ of ghost constraints with poissonian probability
$P_n(h)=e^{-h}h^n/n! $. We can represent this poissonian process in a way that
enables us to use (\ref{eq:derp}) as follows: we assume that each site has
$M>>h$ \emph{slots} connecting it to the background, and that each slot is
occupied by \emph{one} ghost edge with probability $p_h=h/M$. For large $M$,
the probability to have $n$ ghost edges on a site is $P_n(h)$, so one has a
realization of the poissonian distribution. This equivalent system has a total
of $NM$ \emph{ghost slots}, each occupied by a ghost bond with a small
probability $p_h$, so we can now use \Eqn{eq:derp}.
\\
Eliminating a present ghost edge reduces the number of redundant constraints
by one if both of the following conditions are satisfied: {\bf a)} the
remaining constraints that this site has are enough to rigidize (i.e.\ 
$k_\coordn+k_h\geq g$), and {\bf b)} the remaining ghost-constraints that this
site has are not enough to rigidize on their own (i.e.\ $k_h<g$). The joint
probability for these two conditions to be satisfied is
\begin{eqnarray}
&&\sum_{j=0}^{g-1} P_j(h)
\sum_{k=g-j}^{\infty} P_k(x) = \nonumber \\
&&\sum_{j=0}^{g-1} P_j(h) -
\sum_{j=0}^{g-1} P_j(h) 
\sum_{k=0}^{g-j-1} P_k(x) = \nonumber \\
&&R(\coordn,h) - R(0,h) = R-R_h, 
\end{eqnarray}
where we have defined $R_h = G_g(h)$ as the density of rigid sites in the
presence of the field $h$, for a graph with no edges ($\coordn=0$).  The total
number of ghost slots is $NM$ and thus the total number of present ghost bonds
which upon removal produce a change in the number of redundant constraints is
on average
\begin{equation}
NM \times \frac{h}{M} \times (R-R_h)
\end{equation}
\Eqn{eq:derp}, with $p_h=h/M$, now implies that
\begin{equation}
\deriv{\langle B_r \rangle}{h} = N(R-R_h).
\end{equation}
Using \hbox{$\langle B_r\rangle _\coordn=\coordn_r N/2$} one then has
\begin{equation}
\grh = 2(R-R_h).
\label{eq:dergrh}
\end{equation}
\subsection{Checking derivatives of $\coordn_r$}
\label{sec:derivatives-coordn_r}
It is a trivial exercise to show that $\coordn_r$, as given by
(\ref{eq:gammarx}) has the right derivatives, i.e.\ satisfies
\begin{eqnarray}
\grg = R^2
\end{eqnarray}
in agreement with (\ref{eq:deriv2}), and
\begin{eqnarray}
\grh &=& 2 \left \{ R - \R_h \right \},
\end{eqnarray}
in agreement with (\ref{eq:dergrh}).  Now since $\geff = g (1-G_{g+1}(h)) - h
(1-G_{g}(h))$ one has that $\partial \geff/\partial h = G_{g}(h)-1 = \Rh -1$.
Therefore the derivatives of $\udof$ (\Eqn{eq:f2}) are
\begin{eqnarray}
\fg &=&  \frac{R^2 -1}{2}
\nonumber
\\
\fh &=&  R-1
\end{eqnarray}
\section{Derivation of Critical Indices}
\label{sec:deriv-crit-indic}
\Eqns{eq:critconda} and (\ref{eq:critcondb}) imply that, for $y \approx y^*$,
\begin{equation}
R=G_{g}(y)\approx R^*+\frac{\Delta y}{\coordn^*}+\order{(\Delta y)^3},
\label{eq:cubic}
\end{equation}
with $\Delta y = y-y^*$, and $y^*=\coordn^* R^*+h^*$.  The coefficient of the
$\order{(\Delta y)^3}$ term is nonzero. Furthermore $\Delta y=\Delta h +
\coordn^* \Delta R + R \Delta \coordn$, and \Eqn{eq:cubic} can be rewritten as
\begin{eqnarray}
\coordn^* R&=& \coordn^* R^*+ x-x^* + \Delta h+\order{(\Delta y)^3} \qquad
\Rightarrow 
\nonumber
\\ 
R \Delta \coordn  + \Delta h &=& \order{(\Delta y)^3}.
\label{eq:cubic3}
\end{eqnarray}
On the critical isotherm ($\Delta \coordn=0$) we thus have that $\Delta h \sim
(\coordn^* \Delta R + \delta h)^3 \Rightarrow \Delta R \approx (\Delta
h)^{1/3}$, so $\delta=3$.
\\
When $h<h^*$ the order parameter develops a $\coordn$-driven discontinuity
$\Delta R \propto (\Delta h)^{\beta}$, which can be estimated in the following
way. Assume $R^*$ is one of the three solutions of \Eqn{eq:cubic} when $\Delta
h<0$, and determine the value of $\coordn$ for which this happens. Because
$\Delta y=0$ and $\Delta R=0$, one finds that $\Delta \coordn = -1/R^* \Delta
h$. Plugging this result back into \Eqn{eq:cubic} and eliminating the solution
$R=R^*$ one finds that the other two solutions behave as $\Delta R \propto
(\Delta h)^{1/2}$, or $\beta =1/2$.
\\
The ``susceptibility'' $\chi = \deriv{R}{h}|_{h^*}$ diverges on approach to the
critical point as $\chi \propto (\Delta \coordn)^{-\gamma}$. Deriving
\Eqn{eq:h2} one gets
\begin{equation}
\deriv{R}{h} = \frac{P_{g-1}}{1-\coordn P_{g-1}},
\end{equation}
and recalling that $\coordn^*=(P_{g-1}(y^*))^{-1}$ we find that $\chi \propto
(\Delta \coordn)^{-1}$, or $\gamma=1$.
\\
Notice that Rushbrooke's ($\alpha + 2\beta + \gamma=2$) and Griffith's
($\alpha+\beta(\delta+1))=2$) relations are satisfied with $\alpha=0$, and
this is consistent with the fact that the constant-$R$ specific heat
(\ref{eq:Cv}) is zero.

\end{document}